\documentclass[a4paper,12pt]{article}
\usepackage{soul}
\usepackage[usenames,dvipsnames]{color}
\usepackage{jheppub}
\usepackage{esvect}
\usepackage{amsmath, amssymb, slashed, epsf, color, graphicx, latexsym}
\usepackage{epsfig}
\usepackage{graphics}

\newcommand{\BR}{{\bar{R}}}
\newcommand{\Bnabla}{{\bar{\nabla}}}

\newcommand{\ph}{\phi}          
\newcommand{\ps}{\psi}

\begin{document}
	
	\title{Large $D$ Black Holes in an environment}
	\author[a]{Taniya Mandal}
	\author[b] {, Arunabha Saha}
	\affiliation[a]{School of Physics and Mandelstam Institute for Theoretical Physics, University of the Witwatersrand, Wits, 2050, South Africa} 
	\affiliation[b]{University of Geneva, 24 quai Ernest-Ansermet, 1211 Geneve 4, Switzerland}
	\emailAdd{taniya.mandal@wits.ac.za}
	\emailAdd{arunabha.saha@unige.ch}
	
\abstract{We construct dynamical black hole solutions to Einstein Equations in presence of matter in the large $D$ limit. The matter stress tensors that we consider are weak in the sense that they source asymptotic spacetimes with internal curvatures of the order of $\mathcal{O}(D^0)$. Apart from this, we work with a generic stress tensor demanding only that the stress tensor satisfies the conservation equations. The black hole solutions are obtained in terms of the dual non-gravitational picture of membranes propagating in spacetimes equivalent to the asymptotes of the black holes. We obtain the metric solutions to the second sub-leading order in $1/D$. We also obtain the equations governing the dual membranes up to the first sub-leading order in $1/D$. }

\maketitle
\section{Introduction}
The large $D$ regime where one takes the dimensionality of the spacetime $D$ to tend to infinity has been found to be a very useful limit to study black hole dynamics. Dynamical black holes in this limit are found to be dual to the non-gravitational systems of the co-dimension one membranes propagating in spacetimes with the same metric as the asymptotes of the black holes.~This is the so called `Large $D$ Black Hole Membrane Paradigm'.~This duality was established in \cite{Bhattacharyya:2015dva} at leading non-trivial order in $1/D$ for asymptotically flat black hole solutions of vacuum Einstein Equations. Subsequently the duality has been established at sub-leading order in $1/D$ \cite{Dandekar:2016fvw}, for charged black holes \cite{Bhattacharyya:2015fdk}, for black hole with cosmological constant \cite{Bhattacharyya:2017hpj, Bhattacharyya:2018szu}, for charged black hole with cosmological constant \cite{Kundu:2018dvx} and for higher derivative theories of gravity \cite{Saha:2018elg, Kar:2019kyz, Dandekar:2019hyc}. \footnote{In a different approach the effective non-gravitational system dual to a large $D$ black hole was obtained in terms of the local mass and momentum variables of an object moving in the asymptotic spacetime of the black hole \cite{Emparan:2015hwa, Suzuki:2015iha}. In \cite{Dandekar:2016jrp} an equivalence between these two approaches was established for black branes dynamics. In addition, the spectrum of linearised excitations about effective systems dual to static black holes in both these approaches produces the correct light quasi-normal modes spectrum of the static black holes\cite{Emparan:2014aba, Emparan:2015rva}. No general proof of equivalence between these two approaches has been established yet.}

In this paper, we establish the large $D$ membrane paradigm for dynamical black hole solutions of Einstein equations sourced by  `weak' matter stress tensors. We call these matter stress tensors to be weak because they support only those asymptotic spacetimes which have Riemann curvature tensors of order $\mathcal{O}(D^0)$ which turn out to be $\mathcal{O}(D^2)$ orders smaller than the curvature in the regions warped by the black holes. We find that the membranes dual to these black holes move in spacetimes equivalent to the asymptotes of the black hole solutions. We also find the equations governing the dynamics of these dual membranes. 

The broader goal of this project is to understand the properties of the membranes dual to black hole solutions in the most generic situation, e.g., in \cite{Saha:2020zho} it was established that the general structures of the dual membrane equations for black holes in arbitrary higher derivative theories of gravity are constrained by the second law of thermodynamics. By studying black holes in presence of generic matter fields we hope to uncover constraints on matter stress tensors (beyond the null energy conditions). The present work is an initial excursion in this direction where we study the effect of a very weak stress tensor for the Einstein-Hilbert theory of gravity on the membrane equations. 

Due to the weakness of the matter stress tensor non-trivial effects on the membrane equations only appear at the first sub-leading order in $1/D$. The membrane equations that we obtain in this paper are given by

\begin{equation}\label{eq:constraint22}
\begin{split}
&\left[\frac{\hat{\nabla}^2 u_A}{\cal{K}}-\frac{\hat{\nabla}_A{\cal{K}}}{\cal{K}}+u^B {\cal{K}}_{BA}-u\cdot\hat{\nabla} u_A\right]{\cal P}^A_C+ \Bigg[-\frac{u^B {\cal{K}}_{B D} {\cal{K}}^D_A}{\cal{K}}+\frac{\hat{\nabla}^2\hat{\nabla}^2 u_A}{{\cal{K}}^3}-\frac{(\hat{\nabla}_A{\cal{K}})(u\cdot\hat{\nabla}{\cal{K}})}{{\cal{K}}^3}\\
&-\frac{(\hat{\nabla}_B{\cal{K}})(\hat{\nabla}^B u_A)}{{\cal{K}}^2}-\frac{2{\cal{K}}^{D B}\hat{\nabla}_D\hat{\nabla}_B u_A}{K^2} -\frac{\hat{\nabla}_A\hat{\nabla}^2{\cal{K}}}{{\cal{K}}^3}+\frac{\hat{\nabla}_A({\cal{K}}_{BD} {\cal{K}}^{BD} {\cal{K}})}{K^3}+3\frac{(u\cdot {\cal{K}}\cdot u)(u\cdot\hat{\nabla} u_A)}{{\cal{K}}}\\
&-3\frac{(u\cdot {\cal{K}}\cdot u)(u^B {\cal{K}}_{BA})}{{\cal{K}}}
-6\frac{(u\cdot\hat{\nabla}{\cal{K}})(u\cdot\hat{\nabla} u_A)}{{\cal{K}}^2}+6\frac{(u\cdot\hat{\nabla}{\cal{K}})(u^B {\cal{K}}_{BA})}{{\cal{K}}^2}+3\frac{u\cdot\hat{\nabla} u_A}{D-3}\\
&-3\frac{u^B {\cal{K}}_{BA}}{D-3}\Bigg]{\cal P}^A_C +\frac{T^c_{AB} u^B}{\cal{K}}{\cal P}^A_C = {\cal{O}}\left(\frac{1}{D}\right)^2,
\\
\\
&\hat{\nabla}\cdot u = \frac{1}{2{\mathcal K}}\left( \hat{\nabla}_{(A}u_{B)}\hat{\nabla}_{(C}u_{D)}{\cal P}^{BC}{\cal P}^{AD} \right)+ {\cal{O}}\left(\frac{1}{D}\right)^2. 
\end{split}
\end{equation} 
Here, the $A, B,\ldots$ indices run over the $D-1$ directions on the membrane world-volume for the membrane propagating in the target space. In the above equation, $\hat{\nabla}$ is the covariant derivative w.r.t  the induced metric on the membrane world-volume. $\mathcal{K}_{AB}$ is the extrinsic curvature of the membrane embedded in this spacetime. $\mathcal{P}^A_C$ is the projector orthogonal to the velocity vector in the membrane world-volume. From the form of the membrane equations, it is clear that the effect of the asymptotic stress tensor on the membrane equations is two-fold. The vector membrane has an explicit dependence on the asymptotic stress tensor. In addition to that, the stress tensor has the implicit effect of changing the asymptotic spacetime and hence the induced metric in the membrane world-volume. 

We also obtain the metric of the dual black holes in terms of the membrane shape and velocity, the curvature tensors of the asymptotic spacetimes and the matter stress tensors. In the next section, we will discuss the conventions that we have used and also elaborate the procedure to take the large $D$ limit making sure that we capture non-trivial physics. 

\section{The effective gravity equations with $SO(D-p-2)$ isometry}
We are interested in dynamical black hole solutions in the large $D$ limit for Einstein-Hilbert theory of gravity with matter stress tensor. The gravity equations are given by
\begin{equation}
\mathcal{R}_{AB}-\frac{1}{2}g_{AB} \mathcal{R}=8\pi \tilde{T}_{AB}=T_{AB},
\end{equation}
where $\tilde{T}_{AB}$ is the matter stress tensor which is a source for the Einstein tensor. For our convenience, we will choose to work with the scaled stress tensor $T_{AB}$ throughout this paper. Following \cite{Bhattacharyya:2015dva, Bhattacharyya:2015fdk} etc., we take the large $D$ limit while allowing for only those dynamics that preserve a $SO(D-p-2)$ isometry with $p$ being held fixed and finite i.e. we confine non-trivial dependence on spacetime coordinates only in a fixed and finite number of directions while preserving a large isometry in the $D\rightarrow \infty$ limit. 

 Spacetimes with metrics which have $SO(D-p-2)$ isometry can, without loss of any generality, be written in the form \cite{Bhattacharyya:2015dva}
\begin{equation}
	ds^2=g_{\mu\nu}dx^\mu dx^\nu+e^{\phi(x)}d\Omega_{D-p-3}^2.
\end{equation}
Here, $x^\mu$ denotes the coordinates in the $p+3$ number of directions along which non-trivial dynamics take place. The fields $g_{\mu\nu}$ and $\phi$ depend only on the coordinates $x^\mu$ and hence the full spacetime can be thought of as being parametrised by a metric $g_{\mu\nu}$ of the effective $p+3$ dimensional spacetime and the dilaton field $\phi$ propagating in this effective spacetime. 

A general stress tensor that can support this geometry need to also preserve this isometry. i.e. the components of the stress tensor along the isometry directions must be proportional to the unit sphere metric, i.e.  $$T_{ij}=\bar{T}e^{\phi(x)}\Omega_{ij}.$$
The components of the stress tensor along the effective spacetime directions are denoted by $T_{\mu\nu}$. The stress tensor components (i.e. $T_{\mu\nu}$ and $\bar{T}$) are only functions of the effective spacetime coordinates $x^\mu$ . The $D$ dimensional gravity equations can be expressed as effective gravity equations for the metric and the dilaton field in the $p+3$ dimensional effective spacetime spanned by the $x^\mu$ coordinates given by
\begin{eqnarray}\label{effecive_grav_equation1}
	&&\bar{R}_{\alpha\beta}-\frac{d}{2}\Bnabla_{\alpha}\Bnabla_{\beta}\phi-\frac{d}{4}\Bnabla_{\alpha}\phi\Bnabla_{\beta}\phi-\frac{1}{2}g_{\alpha\beta}\Big(\BR-d\Bnabla^2\ph-\frac{d(d+1)}{4}\Bnabla_{\alpha}\phi\Bnabla^{\alpha}\phi+d(d-1)e^{-\phi}\Big)=T_{\alpha\beta},\nonumber\\&&\nonumber\\
	&&\text{and,}\nonumber\\&&\nonumber\\&&-\frac{(d-1)(d-2)}{2}e^{-\phi}+\frac{d-1}{2}\bar{\nabla}^2\phi+\frac{d(d-1)}{8}\bar{\nabla}_\alpha\phi\bar{\nabla}^\alpha\phi-\frac{\bar{R}}{2}=\bar{T}.
\end{eqnarray}

\section{Equation in Patch coordinate}
One of the interesting properties of asymptotically flat black holes in the large $D$ limit is that the non-trivial warping of spacetime is confined within a very short distance of order $\mathcal{O}(r_h/D)$ outside the horizon of length scale $r_h$. This is easy to see for static black holes with metric 
\begin{equation}
	ds^2=-f(r) dt^2+\frac{dr^2}{f(r)}+r^2 d\Omega_{D-2}^2\quad \text{where,}\quad f(r)=1-\left(\frac{r_0}{r}\right)^{D-3}.
\end{equation}
In the above metric at spacetime points where $r-r_0\sim \mathcal{O}(D^0)$ the metric is effectively flat in the $D\rightarrow\infty$ limit. If we use a new radial coordinate `$R$' given by $$r=r_0\left(1+\frac{R}{D}\right),$$ 
we find that the spacetime has non-trivial warping in regions of spacetime where $R\sim\mathcal{O}(D^0)$ (i.e. $r-r_0\sim\mathcal{O}(D^{-1})$). 
For dynamical black holes, this property turns out to be true as is evident from the quasi-normal mode analysis of \cite{Emparan:2014aba} and also from the subsequent non-linear analysis in the large $D$ membrane paradigm papers. We think of this as the defining property of the black hole solutions which in the large $D$ limit have a dual non-gravitational description in terms of co-dimension one membranes. Hence, we demand that the dynamical black hole solutions in presence of matter stress tensor also have spacetime with a similar structure.  More precisely we assume that the matter stress tensor can support black hole solutions that have internal curvatures of the order of $\mathcal{O}(D^2)$ in the ``membrane region" of $\mathcal{O}(1/D)$ width around the horizon and then it exponentially reaches the asymptote with non-zero internal curvatures of order $\mathcal{O}(D^0)$ at $\mathcal{O}(D^0)$ distance away from the horizon. 

Another interesting observation of the QNMs of large $D$ black holes of \cite{Emparan:2014aba, Emparan:2015rva} is that the light quasi-normal modes (the long-lived modes) are the only modes whose support is confined in the region of width $\mathcal{O}(r_h/D)$ outside the horizon. This extends in the non-linear regime where the dynamical black hole solutions for which the position of the horizon and the generator of the event horizon are functions of spacetime with derivatives of the order $\mathcal{O}(D^0)$ have non-trivial physics confined at a distance $\mathcal{O}(r_h/D)$ outside the horizon.~The $D$ dependence of the curvatures of the spacetime in the region outside the horizon is due to the explicit $D$ dependent power-law dependence of the metric on the distance from the horizon.~In \cite{Bhattacharyya:2015dva, Bhattacharyya:2015fdk} it was shown that a very convenient way to capture this phenomenon is to zoom into any arbitrary point $x_0$ in the membrane region and work with a new set of coordinates based around $x_0$ such that this coordinate chart only covers the region up to the end of the membrane region where the spacetime reaches the asymptote. Since the width of the membrane region is $\mathcal{O}(1/D)$, this new coordinate system can be represented by the coordinates $y^a$ given by
\begin{eqnarray}\label{coord}
x^\mu=x^\mu_0+\frac{1}{D-3}~ \alpha^{\mu}_a y^a.
\end{eqnarray}	
From the gravity equations in the effective space-time, it is clear that the metric and the dilaton have to vary at different length scales for them to contribute to the equations at the same order. 
In \cite{Bhattacharyya:2015dva, Bhattacharyya:2015fdk} it was shown that by appropriately rescaling the fields and their derivatives it is possible to confine oneself to those dynamics where every term in the equation of motion contributes at the same order in $1/D$ at the leading order in large $D$ limit. We briefly review the procedure to arrive at the right scaling for the metric and the dilaton field below. 
 
 Under the coordinate transformation from $x^\mu$ to the new coordinates $y^a$ the fields transform as
 \begin{eqnarray}
 	&&g_{ab}=\frac{\partial x^\mu}{\partial y^a}\frac{\partial x^\nu}{\partial y^b}g_{\mu\nu}=\frac{1}{(D-3)^2}\alpha^\mu_a\alpha^\nu_bg_{\mu\nu},\nonumber\\
 	&&\partial_a\phi=\frac{1}{D-3}\alpha^\mu_a\partial_\mu\phi.\nonumber
 \end{eqnarray}
 The rescaling necessary to make sure that the terms in the gravity equation contribute at the same order in $1/D$  are given by
 \begin{eqnarray}
 	&&G_{ab}=(D-3)^2g_{ab}=\alpha^\mu_a\alpha^\nu_bg_{\mu\nu},\nonumber\\
 	&&G^{ab}=\frac{1}{(D-3)^2}g^{ab}=\alpha^a_\mu\alpha^b_\nu g^{\mu\nu},\nonumber\\
 	&&\chi_a=(D-3)\partial_a\phi=\alpha^\mu_a\partial_\mu\phi.\nonumber
 \end{eqnarray}
The rescaled metric $G_{ab}$ and the field $\chi_a$ associated with the dilaton are the fundamental fields in terms of which we write down the gravity equations in the new $y^a$ coordinates. It is easy to see that the rescaling of the stress tensor which makes sure that they contribute to the gravity equation at the same order as the other fields is given by
\begin{eqnarray}\label{stresst_tensor_scaling}
	&&T^{(g)}_{ab}\rightarrow T^{(G)}_{ab},\nonumber\\
	&&\bar{T}^{(g)}\rightarrow (D-3)^2\bar{T}^{(G)}.
\end{eqnarray}
Under the scaling mentioned above  the effective gravity equations in the patch coordinates $y^a$ are given by
\begin{eqnarray}
	&&\bar{ R}_{ab}-\frac{d\epsilon}{2}\bar{\nabla}_a\chi_b
	-\frac{d\epsilon^2}{4}\chi_a\chi_b - \frac{G_{ab}}{2}\bigg(\bar{R}-d\epsilon \bar{\nabla}_a\chi^a -\frac{d(d+1)\epsilon^2}{4}\chi_a\chi^a+\frac{d(d-1)\epsilon^2}{\sigma^2}\bigg)- T_{ab} = 0,\nonumber\\
	&& -\frac{(d-1)(d-2)}{2\sigma^2}-\frac{d-1}{2\epsilon}\bar{\nabla}_a\chi^a-\frac{d(d+3)}{8}\chi_a\chi^a-\frac{\bar{R}}{2\epsilon^2}-\frac{\bar{T}}{\epsilon^2} = 0.\nonumber
\end{eqnarray}
We show in appendix \ref{Emax}, that the matter stress tensor arising out of $U(1)$ gauge fields of the Einstein-Maxwell theory follow the scaling of \eqref{stresst_tensor_scaling}. Also, we show that the stress tensor due to the cosmological constant does not satisfy these scaling (due to the scaling relations already assumed for the metric fields) if we also do not simultaneously scale the cosmological constant.

As we have mentioned earlier, we will not be working with a particular example of a matter stress tensor. Rather we will investigate the general effects of including a matter stress tensor on the large $D$ membrane paradigm. The only constraint that we put on the general matter stress tensor that we work with is that it satisfies the conservation equations. This is necessary for the stress tensor to consistently couple with the metric via the gravity equations.

\subsection{Conservation of stress tensor}
We now write down the equations governing the conservation of the matter stress tensor in the patch coordinates of the effective spacetime, taking into account the appropriate scaling of the metric and matter fields. The conservation equation of the stress tensor has one free index which can be either in the effective spacetime directions or the isometry sphere directions. It is easy to check that the conservation equations along the isometry sphere directions are satisfied trivially by construction 
\begin{eqnarray}
	\nabla_A T^{Ai}&=&\nabla^{(s)}_j T^{ji} \quad (\because T^{\mu i}=0 ),\nonumber\\
	&=& e^{-\phi}\Omega^{ij}\partial_j \bar{T},\nonumber\\
	&=&0,
\end{eqnarray}
where, $\nabla^{(s)}$ denotes covariant derivatives w.r.t the unit sphere metric along the isometry directions. On the other hand the conservation of stress tensor along effective spacetime directions  reduces to
\begin{eqnarray}\label{cons1}
	\nabla_A T^{A\nu}=\nabla_\mu T^{\mu\nu}+\nabla_i T^{i\mu} &=&0,\nonumber\\
	\implies\bar{ \nabla}_\mu T^{\mu\nu}+\Gamma^i_{i\alpha} T^{\alpha \nu}+\Gamma^\nu_{ij}T^{ij}&=&0,\nonumber\\
	\implies \bar{ \nabla}_\mu \bar{T}^{\mu\nu}+\frac{d}{2}\partial_\mu \phi \bar{T}^{\mu\nu}-\frac{d \bar{T}}{2}\partial^\nu \phi &=& 0.
\end{eqnarray}
Here, $\bar{\nabla}$ denotes the covariant derivative w.r.t the metric of effective spacetime. 
The last line above is obtained by evaluating the Christoffel symbols on the effective spacetime metric. Written in terms of the patch coordinates and the appropriately scaled metric, dilaton and matter fields the conservation equations are
\begin{eqnarray}
	\bar{\nabla}_a \bar{T}^a_b+\frac{d \epsilon}{2} \chi_a \bar{T}^a_b-\frac{d \epsilon \bar{T}}{2}\chi_b =0.
\end{eqnarray}
It is easy to see from the above equation that the scaling of the matter stress tensor that we assumed is appropriate since  every term in the conservation equation after scaling contributes at the same order in $1/D$  at the leading order in the large $D$ limit. 
\subsection{The limit of weak stress tensor}
 We have already stated that the stress tensor that we are interested in are weak, by this we mean that
the asymptotic stress tensor sources internal curvatures of order $\mathcal{O}(D^0)$ in the asymptotic region. Also, we assume that the stress tensor configuration supports a spacetime with $SO(D-p-2)$ isometry. In appendix \ref{weakstress}, we will demonstrate that for a configuration with the large isometry structure that we are concerned about, the stress tensors which are $\mathcal{O}(D^0)$ in the global coordinates give rise to internal curvatures in the asymptote which are also $\mathcal{O}(D^0)$. In the patch coordinates the equivalent statement is that stress tensors of order $\mathcal{O}(\epsilon^2)$ will source internal curvatures of order $\mathcal{O}(\epsilon^2)$ where,  $1/D=\epsilon$ i.e.
\begin{equation}
	T_{ab}^{(G)}\sim \mathcal{O}(\epsilon^2) \quad \text{and,}\quad \bar{T}\sim\mathcal{O}(\epsilon^2).
\end{equation} 
We also assume that the strength of the full stress tensor (in patch coordinates) is of the order $\mathcal{O}(\epsilon^2)$ in the membrane region and that the stress tensors in the two regions smoothly match each other at the asymptote of the membrane region. This behaviour is different from that of the stress tensor of asymptotically flat black holes of the Einstein-Maxwell theory. For the Einstein-Maxwell case, due to the confinement of the warping of the Maxwell fields to the membrane region, the stress tensor in the membrane region is $\mathcal{O}(D^2)$, whereas it is effectively zero outside the membrane region. 

The model of the matter stress tensor that we work with is similar to the stress tensor sourced by the cosmological constant. This stress tensor for cosmological constant is proportional to the metric of the region under consideration. Since the metric fields are $\mathcal{O}(D^0)$ everywhere, the strength of the stress tensor in different regions is the same and can be tuned by tuning the cosmological constant. Another example of this type of stress tensor is the stress tensor of a fluid (see appendix \ref{Emax} for details.). 

One of the reasons that the stress tensor of the cosmological constant and the fluid are more aligned to the model of stress tensor that we are working with is due to the fact that for these stress tensors the conservation of the stress tensor is all the information that we have about the equation of motion of the matter content (modulo the equation of state), whereas for Maxwell like fields the field equations carry more information than the conservation equation of the matter stress tensor. In our model of the matter stress tensor, the on-shell condition for matter is solely given by the conservation equation of the stress tensor. 

Having set up the conventions we now dive into the details of the procedure that we use to arrive at dynamical black hole solutions of the gravity equations with weak matter stress tensor. 

\section{Leading order  ansatz solution for the gravity equation}
To find a black hole solution in a $1/D$ expansion we need to figure out a suitable leading order ansatz black hole metric that satisfies the gravity equations to the leading non-trivial order in $1/D$. Since the general matter stress tensor that we are working with has characteristics similar to the stress tensor of the cosmological constant ( up to overall $D$ dependent factors), we briefly review below the mechanism to arrive at the ansatz metric for black holes which asymptote to solutions of gravity equations with the cosmological constant. The analysis of the next section is based on discussions in \cite{Bhattacharyya:2017hpj, Bhattacharyya:2018szu} with conventions used in \cite{Kar:2019kyz}. 
\subsection{Black Holes in asymptotic $AdS$ spacetime}
Static Black holes which are asymptotically $AdS$ have the following metric in the ``Kerr-Schild" coordinate system
\begin{equation}
	ds^2=\boxed{-(1+r^2 \lambda) dt^2+\frac{dr^2}{1+r^2 \lambda}+r^2 d\Omega^2_{D-2}}+\left(\frac{r_h}{r}\right)^{D-3}\left(\sqrt{1+r^2\lambda}dt+\frac{dr}{\sqrt{1+r^2\lambda}}\right)^2.
\end{equation}
These are solutions of the gravity equations in presence of a cosmological constant,
\begin{equation}
	R_{AB}-\frac{1}{2}R g_{AB}=\frac{(D-1)(D-2)}{2}\lambda g_{AB}.
\end{equation}
The factors of $D$ on the RHS above ensure that there is no explicit $D$ dependence in the metric of the asymptotic $AdS$ spacetime.~The RHS of the above equation is the stress tensor due to the cosmological constant.~The stress tensor at any point in the spacetime due to the cosmological constant is proportional to the metric field at that point in the spacetime.~The Kerr-Schild metric written above is a combination of an $AdS$ part (which is written inside a box above ) and a part that parametrises the warping of the spacetime away from the horizon. This warping factor characterises the deviation of the spacetime from the local $AdS$ metric at any point in the spacetime. 
The Kerr-Schild metric can be written in a covariant form as
\begin{eqnarray}
	ds^2_{ansatz}=ds^2_{asym}+\frac{(O_M dx^M)^2}{\psi^{D-3}},
\end{eqnarray}
where,
$$\psi=\frac{r}{r_h},\quad O_M=n_M- u_M\quad \text{where}\quad n_M=\frac{dr}{\sqrt{dr\cdot dr}}\quad \text{and,}\quad u_M=-\frac{dt}{\sqrt{dt\cdot dt}}.$$
The $ds^2_{asym}$ part is the $AdS$ metric written inside the box of the Kerr-Schild form of the metric and the dot products are taken w.r.t the $ds_{asym}^2$ metric. In \cite{Bhattacharyya:2017hpj}, the Kerr-Schild form of the metric was used as a starting point to arrive at the staring ansatz metric for the dynamical black hole which solves for the gravity equations to leading order. This ansatz metric is obtained by elevating $\psi$ and $u_M$ to be functions of spacetime coordinates with $\mathcal{O}(D^0)$ derivatives in the global coordinates. It was shown that the ansatz metric then solves the gravity equations at leading order in $1/D$ if 
\begin{equation}
	\nabla^2\left(\psi^{-(D-3)}\right)=\mathcal{O}(D^0)\quad \text{and,}\quad \nabla\cdot u=\mathcal{O}(D^0).
\end{equation}
The black hole asymptotes to the metric $ds_{asym}^2$ as  $\psi-1\sim\mathcal{O}(D^0)$ (or $r-r_h\sim r_h \mathcal{O}(D^0)$ for the static black hole), in the large $D$ limit. The interesting thing about the structure of the gravity equations and the ansatz metric solution is that even in the membrane region, $ds_{asym}^2$ is a solution to the gravity equations with stress tensor sourced by the cosmological constant for the metric of the asymptotic region but evaluated at the point of interest in the membrane region. We take this as a model to construct the ansatz metric for more general matter stress tensor as we explain next.  

\subsection{Ansatz metric in presence of a general matter stress tensor}

Let us focus on a small region of the length scale of the order of $\mathcal{O}(r_h/D)$ around an arbitrary point in the membrane region $x_0$. We assume that the ansatz metric in presence of a general matter stress tensor will also be a sum of two parts: $1)$ The $ds_{asym}^2$ part which denotes the asymptotic metric of the black hole but evaluated at $x_0$ and $2)$ the part characterising the non-trivial  warping of the spacetime as we go away from the horizon, i.e. 
\begin{equation}
	ds_{ansatz}^2=ds_{asym}^2+ds_{warp}^2(\psi).
\end{equation}
 Let the matter stress tensor sourcing the full ansatz metric be denoted by $T_{MN}(x)$ and let the component of stress tensor asymptote to $T_{MN}^{c}$ at $x_{asym}$ which is a point near the asymptote of the membrane region. The metric of the spacetime at $x_{asym}$ is effectively equal to $ds_{asym}^2$. We know that at $x_{asym}$ the metric $ds_{asym}^2$ solves the gravity equations with the matter stress tensor given by $T_{MN}^{c}(x_{asym})$. The matter stress tensor which solves for the $ds_{asym}^2$ at $x_0$ in the membrane region is related to $T_{MN}^{c}(x_{asym})$ by a Taylor expansion
\begin{equation}
	T_{MN}^{c}(x_0)-T_{MN}^{c}(x_{asym})=(x_{asym}^P-x_0^P)\partial_P T_{MN}(x_{asym})+\ldots
\end{equation}
The RHS of the above equation starts at $\mathcal{O}(1/D)$ times the leading order value of the stress tensor, since $x_{asym}-x_0\sim \mathcal{O}(1/D)$ and the derivatives of the asymptotic stress tensor are the same order as the stress tensor. Hence, the stress tensor which parametrises the $ds_{asym}^2$ part of the ansatz metric in the membrane region is effectively equivalent to the boundary value of the matter stress tensor of the membrane region to leading order in $1/D$.

The next question that arises is how do we unambiguously define the point $x_{asym}$ that corresponds to the asymptotic counter-part of $x_0$.  More precisely, given a point $x_0$ in the membrane region, which of the points in the asymptotic region which are $\mathcal{O}(1/D)$ distance away from $x_0$ do we use to evaluate the boundary value of the membrane stress tensor and hence effectively the stress tensor sourcing the $ds_{asym}^2$ part of the ansatz metric.~In order to resolve this let us consider another point in the asymptote, say $x_{asym}^{(2)}$ which is also at $\mathcal{O}(1/D)$ distance from $x_0$. The above argument holds for this new point too and one concludes that the stress tensor at $x_{asym}^{(2)}$ is also a good candidate for the stress tensor for $ds_{asym}^2$ at $x_0$. Also the difference in value of the asymptotic stress tensor between the points $x_{asym}$ and $x_{asym}^{(2)}$ is also $\mathcal{O}(1/D)$ since $x_{asym}-x_{asym}^{(2)}\sim\mathcal{O}(1/D)$ for $x_0-x_{asym}^{(2)}$ to be of the order $\mathcal{O}(1/D)$ and hence, we can choose either point as the candidate for the asymptotic counter-part of $x_0$.  We choose the point that we reach from $x_0$ if we move along $d\psi$ from $x_0$, as a convenient candidate for $x_{asym}$. Hence, the stress tensor sourcing the $ds_{aym}^2$ part of the ansatz metric in the membrane region is the boundary value of the stress tensor in the membrane region if we move along the $d\psi$ direction away from the horizon\footnote{The analysis for the correct value of the asymptotic stress tensor should in principle have been done in terms of the Lie derivatives of the stress tensor along the direction to the asymptotic point. This does not make any difference to our conclusion since this change would still have led to a RHS of order $\mathcal{O}(1/D)$ in the expression for the difference of the asymptotic stress tensor at $x_0$ and $x_{asym}$}.

One more piece of information that we have is that the asymptotic stress tensor is such that it sources internal curvatures of order $\mathcal{O}(D^0)$. Hence, the Taylor series expansion of the $ds_{asym}^2$ part of the ansatz metric in a region of length scale $\mathcal{O}(1/D)$ around $x_0$ resembles flat space to leading order in $1/D$ and has corrections at subsequent orders in $1/D$. In addition, the full stress tensor in the membrane region is weak (i.e it is of order $\mathcal{O}(\epsilon^2)$ in patch coordinates). Thus, to leading order around the point $x_0$, the ansatz metric is practically a solution of vacuum Einstein equations which asymptote to flat spacetime. Hence, to leading order the ansatz metric of the black hole solutions with weak matter stress tensor is 
\begin{equation}
	ds_{asym}^2+\frac{(O_M dx^M)^2}{\psi^{D-3}},
\end{equation}
where $O_M$ and $\psi$ are $\mathcal{O}(D^0)$ derivative functions of the spacetime coordinates constructed in an identical fashion as in the case of gravity equations with cosmological constant. The only difference is that the normalisation properties of the normal and velocity vector fields on the membrane are set w.r.t the asymptotic metric $ds_{asym}^2$. The ansatz metric solves for the gravity equations to the leading order if the following two conditions are satisfied on the functions $\psi$ and $u_M dx^M$
\begin{equation}
	\nabla^2\left(\frac{1}{\psi^{D-3}}\right)=\mathcal{O}(D^0)\quad \text{and,}\quad \nabla\cdot u=\mathcal{O}(D^0).
\end{equation}
These are the same conditions that the shape and velocity function need to satisfy for the ansatz metric for black holes in asymptotic flat and $AdS$ space as well. It is easy to see that these conditions result in the solution to the gravity equations because to the leading ( and first sub-leading) order the metric and the gravity equations are effectively equivalent to those of black holes in asymptotic flat spacetime. 
\section{Parametrising components of the ansatz metric}
In the last sub-section, we have seen that the isometry of the problem and the weak nature of the matter stress tensor in the large $D$ limit allow us to arrive at a relatively simple ansatz metric that looks very similar to the ansatz metric for gravity equations in vacuum and in presence of cosmological constant. In this sub-section, we describe the details of a suitable parametrisation that we have used for the ansatz metric.~To start with we describe the parametrisation of $ds_{asym}^2$ in terms of the unconstrained components of the asymptotic stress tensor and the metric field. 
\subsection{Parametrisation of the asymptotic matter stress tensor}

The $ds_{asym}^2$ part of the ansatz metric about any arbitrary point $x_0$ in the membrane region is obtained by solving the gravity equations with a matter stress tensor which is effectively equivalent to the boundary value of the full matter stress tensor of the membrane region. 

 We look for solutions where, in the membrane region (in global coordinates), the derivatives of the metric and the matter fields along $d\psi$ can at the most be of order $\mathcal{O}(D)$ whereas the other derivatives are $\mathcal{O}(D^0)$ or lower. These solutions are therefore weakly equilibrated since the derivatives along time like directions are also effectively zero to leading order (w.r.t the derivatives along $d\psi$). We assume that under this weak equilibration, the boundary value of the stress tensor sourcing the solution $ds_{asym}^2$ is held constant. This can be achieved if we assume the presence of a constant external supply of matter at the appropriate rate to keep the boundary value of the matter stress tensor effectively fixed at $T_{MN}(x_{asym})=T_{MN}^{(c)}$. The derivatives of the stress tensor orthogonal to $d\psi$ direction are not relevant for the analysis at the order in $1/D$ that we are working with due to the fact that the stress tensor is weak. 
\subsection{Parametrisation of $ds_{asym}^2$}
Since, the solutions that we are after have a $SO(D-p-2)$ isometry, the metrics can be expressed in terms of the metric and dilaton field in the effective $p+3$ dimensional spacetime. Since, the metric $ds_{asym}^2$ has $\mathcal{O}(D^0)$ derivatives in the global coordinates about $x_0$, in the patch coordinates these derivatives are $\mathcal{O}(\epsilon)$. We find it convenient to parametrise $ds_{asym}^2$ in the patch coordinates using the Riemann normal coordinates around $x_0$. It can be shown that from the requirements of the $SO(D-p-2)$ isometry and the fact that the maximum order of derivatives acting on $ds_{asym}^2$ is $\mathcal{O}(D^0)$ in global coordinates, the effective $p+3$ dimensional metric and dilaton field parametrising $ds_{asym}^2$ can be written as (for details look at appendix \ref{weakstress})
\begin{eqnarray}
		g^{asym}_{\mu\nu}(x)=\eta_{\mu\nu}+\frac{1}{3D^2}R_{\mu\beta\alpha\nu}|_{x_0}y^{\beta} y^{\alpha}+\ldots \quad \text{and,}\quad \phi=2 \log(S).
\end{eqnarray}
where, $R_{\mu\alpha\nu\beta}$ is the Riemann tensor of $g_{\mu\nu}$ and $S$ is a coordinate in the effective spacetime directions which also denotes the radius of the sphere in the isometry direction. A possible coordinate representation of the $\eta_{\mu\nu}$ part of the above metric is given by
$$\eta_{\mu\nu}dy^{\mu} dy^{\nu}=-dt^2+dr^2+dS^2+\sum_{i=1}^p dx_i dx^i.$$
The terms in the ellipsis in the expression of $g^{asym}_{\mu\nu}$ are dependent on the local values of derivatives of the Riemann tensor. At the order in $1/D$ of concern to us we are agnostic to them. We can see that the effective spacetime metric part of $ds_{asym}^2$ is parametrised by the Riemann normal coordinates of the effective metric itself. This greatly reduces the number of free parameters which parametrise the arbitrary solutions of the gravity equations. 
\subsection{The on-shell parameters of $ds_{asym}^2$}
The parametrisation mentioned in the last two sub-sections is off-shell since we have not imposed the condition on them that they need to solve the gravity equations. As the stress tensor is weak and since we are working in the Riemann normal coordinates, the gravity equations in the patch coordinate to the leading and sub-leading order in $\epsilon$ are trivially satisfied by the metric $ds_{asym}^2$ which is effectively flat at these orders. The first non-trivial effect of the curvatures and the stress tensor start appearing at $\mathcal{O}(\epsilon^2)$ where the gravity equations effectively become a set of algebraic equations in the patch coordinates. A set of independent components of the stress tensor and the Riemann curvature tensors which parametrise the solution to the gravity equations is given by

$$\text{scalar components}\rightarrow T^{c}_{\mu\nu}n^\mu n^\nu, T^{c}_{\mu\nu}n^\mu u^\nu,T^{c}_{\mu\nu}u^\mu u^\nu,$$ $$\text{vector components}\rightarrow T^{c}_{\mu\nu}n^\mu P^\nu_\alpha, T^{c}_{\mu\nu}u^\mu P^\nu_\alpha, $$ $$\text{tensor component}\rightarrow T^{c}_{\mu\nu}P^\mu_\alpha P^\nu_\beta,$$
and the following components of the Riemann tensor at the asymptote
$$\text{scalar component}\rightarrow R_{\alpha\mu\beta\nu}n^\alpha u^\mu n^\beta u^\nu,$$ $$\text{vector components}\rightarrow  R_{\alpha\mu\beta\nu} n^\alpha u^\mu n^\beta P^\nu_\gamma, R_{\alpha\mu\beta\nu} u^\alpha n^\mu u^\beta P^\nu_\gamma, $$$$\text{tensor components}\rightarrow R_{\alpha\mu\beta\nu} n^\alpha  n^\beta P^\mu_\gamma P^\nu_\delta, R_{\alpha\mu\beta\nu} n^\alpha  u^\beta P^\mu_\gamma P^\nu_\delta,$$ 
where $P_{\mu\nu}$ is orthogonal to $n_\mu$, $u_\mu$ and $Z_\mu=\frac{dS_\mu-n_s n_\mu}{\sqrt{1-n_s^2}}$, with $n_s=n.dS$.

\section{Towards dynamical black hole solutions in a $1/D$ expansion}
The above mentioned components of the Riemann curvature and stress tensor are the complete parametrisation of any arbitrary solution of gravity equations which is a suitable candidate for $ds_{asym}^2$. We now have a complete description of the leading order ansatz metric in terms of the shape $\psi$ and velocity $u_M dx^M$ function of the dual membrane propagating in $ds_{asym}^2$. The next task is to add perturbative corrections in $1/D$  to the ansatz metric which solves for the gravity equations to first and second sub-leading order in $1/D$. 

To solve for the metric corrections around any arbitrary point in the membrane region, we choose a convenient coordinate system. Like the earlier papers on the large $D$ membrane paradigm we choose a coordinate system  in the effective spacetime in which three of the coordinates are chosen along the orthogonal directions of $O_\mu dx^\mu$, $n_\mu dx^\mu $ and $Z_\mu dx^\mu=\frac{dS-n_S n}{\sqrt{1-n_S^2}}$ , where $n_S=n\cdot dS$. The rest of the $p$ coordinates in the effective spacetime directions are chosen in the hyperplane orthogonal to these three distinct directions and also orthogonal to each other. The particular choice of these $p$ coordinates orthogonal to the distinct directions in the effective spacetime directions are irrelevant. The patch coordinate $y^a$ based on this coordinate system is given by
\begin{eqnarray}
	&&R=(D-3)(\psi-1),\nonumber\\
	&&V=(D-3)(x^\mu-x_0^\mu)O_\mu,\nonumber\\
	&&z=(D-3)(x^\mu-x_0^\mu)Z_\mu, \quad \nonumber\\
	\text{and} &&y^i=(D-3)(x^\mu-x_0^\mu)Y^i_\mu.
\end{eqnarray}
The above coordinate system is based around a point in the membrane region with coordinates $(\psi=1,x=x_0)$. 
 We now proceed to understand the constraints on the stress tensor that come from the conservation equations in the membrane region. 
\subsection{Conservation of stress tensor in membrane region}
 We know that the derivatives of stress tensor that are relevant are along the $d\psi$ direction only (i.e. the direction in which the metric has non-trivial warping).~The transverse derivatives can be neglected for the computation of metric corrections up to second sub-leading order in $1/D$ as the stress tensor is weak. In the coordinate system that we have chosen to work with in the membrane region, the non-trivial derivatives on the matter stress tensor are only along the $R$ coordinates. Hence, the conservation equations are effectively coupled ordinary differential equations on the stress tensor components.~We have proved earlier that the conservation equations along the isometry directions are trivially satisfied. We classify the conservation equations according to their tensor structure w.r.t the directions orthogonal to the $u$, $n$ and $Z$ vectors in the effective spacetime. The vector component of the conservation equation (i.e. component orthogonal to $n$, $u$ and $Z$ vectors) are given by
\begin{eqnarray}
	&&\frac{\mathcal{K}e^{-R}}{(D-3)}\frac{d}{dR}\Big(\Big(e^{R}{(u.T)}_\mu+(e^R-1){(O.T)}_\mu\Big)P^\mu_\nu\Big)+{\mathcal{C}^v}_\nu(R)=0,\nonumber\\
	&& \text{where}\quad  {\mathcal{C}^v}_\nu(R)=\frac{{(Z.T)}_{\mu}}{S}P^\mu_\nu  .
\end{eqnarray}
 $P^\mu_\nu$ is the projector orthogonal to $n, u$ and $Z$ in the effective spacetime w.r.t $ds_{asym}^2$. The scalar component of the conservation equation along $Z$ direction is given by
\begin{equation}
	\frac{\mathcal{K}e^{-R}}{(D-3)}\frac{d}{dR}\Big(e^{R}{(u.T.Z)}+(e^R-1){(O.T.Z)}\Big)+{\mathcal{C}^z}(R)=0,
\end{equation}
where $C^z(R)=\frac{1}{S}(Z.T.Z-(1-n_S^2) \bar{T})$, and $n_S=n.dS$. Similarly the conservation equation along $u$ direction is given by
\begin{equation}
	\frac{\mathcal{K}e^{-R}}{(D-3)}\frac{d}{dR}\Big(e^{R}{(u.T.u)}+(e^R-1){(O.T.u)}\Big)+{\mathcal{C}^u}(R)=0,
\end{equation}
where $C^u(R)=\frac{1}{s}(Z.T.u)$. The equation along the $O$ direction is given by
\begin{equation}
	\frac{\mathcal{K}e^{-R}}{(D-3)}\frac{d}{dR}\Big(e^{R}{(u.T.O)}+(e^R-1){(O.T.O)}\Big)+{\mathcal{C}^O}(R)=0,
\end{equation}
where $C^O(R)=\frac{\mathcal{K}}{D-3}(\frac{Z.T.O}{n_s}+\frac{e^{-R}}{2}(O.T.O)-\bar{T})$. The general solution to all of the above conservation equations is of the form 
\begin{equation}\label{memstresssol1}
	{(u.T)}_\mu (R)= \left(- \frac{e^{-R}(D-3)}{\mathcal{K}}{\int_0}^R e^x \mathcal{C}_\mu (x)dx\right) -(1-e^{-R}){(O.T)}_\mu(R)+Q_\mu,
\end{equation}
where, $Q_\mu$ is an integration constant. By demanding that the stress tensors are regular everywhere in the membrane region and evaluating the above solution at $R=0$ we can fix the integration constant $Q_\mu$ and write the solution as
\begin{equation}\label{memstresssol}
	{(u.T)}_\mu (R)-(u.T)_\mu(0)= \left(- \frac{e^{-R}(D-3)}{\mathcal{K}}{\int_0}^R e^x \mathcal{C}_\mu (x)dx\right) -(1-e^{-R}){(O.T)}_\mu(R) ,
\end{equation}
where $\mathcal{C}_\mu=(\mathcal{C}^O,\mathcal{C}^z,\mathcal{C}^u,\mathcal{C}^v_i)$. Here $i$ denotes the $p$ directions orthogonal to $n$,$u$ and $Z$ vectors in the effective spacetime.  
We have already written down a parametrisation of the stress tensor part determining $ds_{asym}^2$ earlier. We now need to check that the asymptotic values of the full stress tensor sourcing the membrane and metric match with the stress tensor components sourcing $ds_{asym}^2$. This we do by taking the $R\rightarrow\infty$ limit of the above equations which gives us
\begin{eqnarray}
	(n\cdot T(\infty))_\mu-(u.T)_\mu(0)&=&-\frac{D-3}{\mathcal{K}}\lim_{R\rightarrow\infty}e^{-R}\int_0^R e^x C_\mu(x) dx,\nonumber\\
	&=& -\frac{D-3}{\mathcal{K}}\lim_{R\rightarrow\infty}\frac{\int_0^R e^x C_\mu(x) dx}{e^R},\nonumber\\
	&=&-\frac{D-3}{\mathcal{K}}\lim_{R\rightarrow\infty}C_\mu(R) \quad (\text{Applying L'Hopital's rule}),\nonumber\\
	&=& -\frac{D-3}{\mathcal{K}}C_\mu(\infty).
\end{eqnarray}
From table \ref{tab:table1} and using the leading order value of $\mathcal{K}$, it is easy to see that $$-\frac{(D-3)}{\mathcal{K}}C_\mu(\infty)=(n.T)^{(c)}_\mu,$$
where, $(n.T)^{(c)}$ is the asymptotic value of the component of the stress tensor along the normal direction. But our boundary conditions demand that $$(n.T)_\mu(\infty)=(n.T)^{(c)}_\mu.$$ 
Hence, for our solution to be consistent we need to have 
\begin{equation}\label{stress_bdy_condn}
	(u.T)_\mu(0)=0.
\end{equation}
For convenience we have listed the asymptotic values of some of the components of the stress tensor and $C_\mu(R)$ functions relevant for this section in table \ref{tab:table1}. 
\begin{table}[h!]
	\begin{center}
		\caption{Asymptotic values}
		\label{tab:table1}
		\begin{tabular}{|l|c|r} 
			\hline
			\textbf{Data} & \textbf{Asymptotic value}\\\hline
			$u.T.Z$ & $-n_s u.T^c.n$ \\\hline
			$O.T.Z$ & $-\frac{n_s(1-n_s^2)}{2-n_s^2}(n.T^c.n+u.T^c.u-T^c_{\mu\nu}P^{\mu\nu})+n_s u.T^c.n$\\\hline
			$Z.T.Z$ & $\frac{1-n_s^2}{2-n_s^2}((1+n_s^2) n.T^c.n-(1-2n_s^2)(u.T^c.u-T^c_{\mu\nu}P^{\mu\nu}))$\\\hline
			$C^z$ & $\frac{n_S^2(1-n_s^2)}{S_0(2-n_s^2)} (n.T^c.n+u.T^c.u-T^c_{\mu\nu}P^{\mu\nu})$\\\hline
			$u.T.u$ & $u.T^c.u$\\\hline
			$O.T.u$ & $ u.T^c.n- u.T^c.u $\\\hline
			$C^u$ & $-\frac{n_s}{S_0} u.T^c.n$ \\\hline
			$O.T.O$ & $n.T^c.n-2 u.T^c.n+u.T^c.u$\\\hline
			$C^O$ & $ \frac{n_s}{S_0}(u.T^c.n-n.T^c.n)$\\\hline
		\end{tabular}
	\end{center}
\end{table}
\subsubsection*{Null Energy Condition}
The component of the above constraint on the membrane stress tensor along the $u$ direction is given by
\begin{equation}
	u\cdot T\cdot u(0)=0.
\end{equation} 
From the form of the ansatz metric, one can check that $u^\mu\partial_{\mu}$ is the generator of the horizon of the black hole. Hence, the above condition can be thought of as the lower bound on the stress tensor components along the generator of the horizon allowed by the null energy condition.

\subsection{Evaluating the first order corrected metric}

The ansatz metric solves for the Einstein equations with matter stress tensor only to the leading order in $1/D$.~To arrive at solutions to the Einstein equations at first sub-leading order and so on, one needs to add corrections to the ansatz metric. Equivalently one can add corrections to the metric and the dilaton field of the effective $p+3$ dimensional spacetime and solve for the effective gravity equations in $p+3$ dimensions. We take the latter approach and the corrections can be written in a schematic form as
\begin{eqnarray}
	g_{\mu\nu}=\sum_{k=0} D^{-k}g_{\mu\nu}^{(k)}\quad \text{and,}\quad \phi=\sum_{k=0}D^{-k}\phi^{(k)}.
\end{eqnarray}
The $k=0$ part of the above expansion correspond to the metric and dilaton field of the ansatz metric and other values of $k$ correspond to higher order corrections in $1/D$.
The metric correction written above has a gauge redundancy corresponding to small diffeomorphism and we fix this with the choice of gauge 
$$g_{\mu\nu}O^\mu=g^{(0)}_{\mu\nu}O^\mu=O_\nu,$$
i.e. the metric corrections do not have components along the null vector $O_\mu dx^\mu$. The leading order in $1/D$ piece of the ansatz metric takes the following form in our choice of patch coordinate system in the membrane region
\begin{eqnarray}\label{ansatz_zero}
	g^{(0)}_{\mu\nu}dy^\mu dy^\nu&=&2\frac{S_0}{n_s}dVdR-(1-e^{-R})dV^2+\frac{dz^2}{1-n_s^2}+dy_i dy^i +\mathcal{O}(1/D),\nonumber \\ 
	\phi^{(0)}&=&2\log(S_0).
\end{eqnarray}
With our choice of gauge fixing the most general form of the correction to the effective metric at any order in $1/D$ is given by
\begin{eqnarray}
	g^{(k)}_{\mu\nu}=&& S^{(k)}_{VV}O_\mu O_\nu+2S^{(k)}_{Vz} O_\mu Z_\nu+S^{(k)}_{tr} P_{\mu\nu}+S^{(k)}_{ZZ} Z_\mu Z_\nu\nonumber\\
	&&+ V^{V(k)}_{(\mu}O_{\nu)} + V^{Z(k)}_{(\mu}Z_{\nu)}+\mathcal{T}_{\mu\nu},
\end{eqnarray}
where, $P_{\mu\nu}$ is the projector orthogonal to $O_\mu, n_\mu$ and $Z_\mu$ in the effective spacetime. and $$\mathcal{T}_{\mu\nu}P^{\mu\nu}=0,$$
where, $\mathcal{T}_{\mu\nu}$ does not have component along $O_\mu, n_\mu$ and $Z_\mu$. The metric corrections at each order are obtained by solving the Einstein equations with the matter stress tensor up to the relevant order. The objects contributing to the ansatz metric i.e. the shape $\psi$ and the velocity functions $u_M dx^M$ are to be expanded in a Taylor series expansion about the point $x_0$ to obtain the ansatz metric in a $1/D$ expansion. The ansatz metric expanded in $1/D$ has a schematic form given by
\begin{equation}
	g_{\mu\nu}^{(0)}=\mathcal{G}_{\mu\nu}^i f_i(R)+\frac{1}{D}\mathcal{G}_{\alpha\mu\nu}^j S_j(R) y^\alpha+\mathcal{O}(1/D^2),
\end{equation}
and there exists a similar expansion for the dilaton field. The $ \mathcal{G}_{\mu\nu}^i, \mathcal{G}_{\alpha\mu\nu}^j $ are composed of the Taylor expansion coefficients of the shape and velocity functions in the effective spacetime evaluated at $x=x_0$. At each order in $1/D$ the pieces formed out of Taylor expansion coefficients are multiplied by non-trivial functions of $R$ coordinate only, namely $f_i(R),S_i(R) $ etc. e.g., one can look at the expression \eqref{ansatz_zero} to find the expression of $\mathcal{G}^i_{\mu\nu} f^i(R) $. Since, the Taylor expansion coefficients are ultra-local functions evaluated at $x=x_0$ the non-trivial dependence on coordinates appear only along $R$ direction. This is because of the non-trivial dependence of the ansatz metric on the shape function $\psi$ is obtained via the following mechanism
$$\psi=\left(1+\frac{R}{D}\right) \quad \text{and,}\quad \lim_{D\rightarrow\infty}\psi^{-D}=e^{-R}+\mathcal{O}(1/D).$$ 
The gravity equations evaluated on the ansatz metric solves for the Einstein equations to leading order on $1/D$. At sub-leading order the action of the gravity equations on the ansatz metric leaves behind an expression which is a collection of non-trivial functions of the $R$ coordinate with the coefficient being combinations of various Taylor expansion coefficients of the shape and velocity functions. Hence, it is obvious that to solve for the gravity equations we need to add corrections to the ansatz metric which are effectively functions of the $R$ coordinates only. Thus the Einstein equations acting on the ansatz metric plus the metric corrections give rise to ordinary differential equations (ODE) on the metric corrections. This is the mechanism that makes the large $D$ expansion so useful as we have converted a set of coupled partial differential equations into a set of coupled ordinary differential equations. 

Finally, we have to remove an ambiguity associated with the Taylor expansion data of the shape and velocity functions away from the $\psi=1$ surface. We are after a duality between the membrane and the black hole. So, the black hole metric at all orders in $1/D$ should be expressible in terms of the properties of the dual membrane whose shape is given by $\psi=1$ and on which resides a velocity field $u_M dx^M$. The membrane moves in the asymptotic spacetime of the black hole. There are no physical data in the membrane picture which correspond to the derivatives of the membrane variables away from the membrane surface. These data are available in the gravity picture because we are able to define the $\psi$ and $u_M$ functions even away from the surface $\psi=1$ where the horizon is located. It can be shown \cite{Bhattacharyya:2015dva, Bhattacharyya:2015fdk} that the physical content of the black hole is independent of data of the normal derivatives of the membrane variables and hence one can set them to any convenient value. We choose to follow the convention of \cite{Bhattacharyya:2015dva} and set the covariant normal derivatives of the normal and velocity fields to be zero, i,.e. 
$$n\cdot\nabla n_M=0\quad \text{and,}\quad n \cdot\nabla u_M=0. $$

The mechanism to evaluate the metric and dilaton field follows this pattern at all orders in $1/D$ too. We only need to make sure that we have an on-shell solution at a particular order in $1/D$ before proceeding to evaluate the corrections at the subsequent order.

\subsection{The Leading order metric corrections and membrane equations}
Since, the effect of the non-trivial aspects of $ds_{asym}^2$ do not appear at the leading and first sub-leading order in $1/D$, we can effectively think of $ds_{asym}^2$ as flat spacetime at this order and hence the results of the metric correction at this order look identical to those obtained in \cite{Dandekar:2016fvw} with all covariant derivatives now being taken w.r.t $ds_{asym}^2$. Hence, the metric and dilaton corrections at leading order in $1/D$ are given by
\begin{eqnarray}
S^{(1)}_{VV} &=& -e^{-R}R^2+\frac{S_0}{n_s}\left(Re^{-R}+\frac{R^2e^{-R}}{2}\right)s_1-\frac{S_0}{n_s^2}R^2e^{-R}s_2,\nonumber\\
S^{(1)}_{Vz} &=& \frac{R e^{-R}S_0}{1-n_s^2}\left(\frac{s_2}{n_s}-s_1\right), \quad
S^{(1)}_{ZZ} = 0,\quad
S^{(1)}_{Tr}=0,\quad
\phi^{(1)} = 0,\nonumber\\
V^{(1)}_{Vi} &=& \frac{Re^{-R}S_0}{n_s}\left(v_{1\alpha}-v_{3\alpha}\right)P^\alpha_i,\quad V^{(1)}_{Zi}= 0,\quad
\mathcal{T}^{(1)}_{\mu\nu} = 0,
\end{eqnarray}
where $s_1=u.\mathcal{K}.u, s_2=u.\mathcal{K}.Z, v_1^\mu=u^\alpha \mathcal{K}_{\alpha\beta}P^{\beta\mu}, v_2^\mu=Z^\alpha \mathcal{K}_{\alpha\beta}P^{\beta\mu}  $ and $P^{\mu\nu}$ is the projector orthogonal to $n$, $u$ and $Z$.

The above solution satisfies all the right boundary conditions on the metric corrections and is regular everywhere in the membrane region. Nevertheless, not all of the Einstein equations are solved yet. It turns out that some of the constraint Einstein equations are solved only when we impose a set of constraints on the membrane data at $\ps=1$. These constraints when expressed in terms of the variables of the membrane world volume are called the membrane equations. 

Before we write down the membrane equations at this order we must mention that the metric and dilaton corrections can be written in a manifestly $p$ independent manner. In \cite{Bhattacharyya:2015fdk} it was shown that the reason why the results are independent of $p$ is that the above result can be expressed in terms of the expression of a metric correction of the black hole written in terms of the full spacetime and demanding that there exists a large isometry in the metric without the need to explicitly mention the number of dimensions in the non-isometric directions. This method was called 'geometrisation' \cite{Bhattacharyya:2015fdk}. We will mention further details of this method in our context in a later section but we present the membrane equations in the geometrised picture here given by

\begin{eqnarray}
	&&\left[\frac{\hat{\nabla}^2 u_A}{\cal{K}}-\frac{\hat{\nabla}_A{\cal{K}}}{\cal{K}}+u^B {K}_{BA}-u\cdot\hat{\nabla} u_A\right]{ \mathcal{P}}^A_C=\mathcal{O}(1/D),\nonumber\\
	&&~~~~~\hat{\nabla}\cdot u=\mathcal{O}(1/D),
\end{eqnarray}
where $A,B,\cdots$ denote the coordinates on co-dimension one membrane. $\mathcal{K}_{AB}$ is the extrinsic curvature of the membrane and $\mathcal{K}$ is the trace of that. The covariant derivatives and the dot products in the above equation are taken w.r.t the induced metric on the membrane world-volume embedded in $ds_{asym}^2$. $\mathcal{P}_{AB}$ is the projector orthogonal to the membrane velocity on the membrane world-volume. Having solved for the metric correction $g_{\mu\nu}^{(1)}$ and dilaton field $\phi^{(1)}$ we now proceed to solve for the metric correction and dilaton field at the second-sub-leading order.

\subsection{The metric corrections at the second sub-leading order}

Since the non-trivial dependence on the curvature of $ds_{asym}^2$ and the stress tensors start appearing at this order we will present the explicit form of the ordinary differential equations that we need to solve in order to arrive at the metric and dilaton corrections in the effective spacetime at this order. The solutions for the metric corrections are subdivided according to their tensor structures w.r.t. the directions orthogonal to the $n_{\mu}$, $u_{\mu}$ and $Z_{\mu}$ in the following sub-sections. 

The stress tensor and the curvatures of the $ds_{asym}^2$ part appear in the ODE determining the gravity equations for the first time at the second-sub-leading order. Hence, they contribute to the gravity equations in a linear manner. It is easy to see that there is no other possibility by considering for example the possibility of coupling of the stress tensor and the curvature contributions.~As each of these objects has an inherent strength of $\mathcal{O}(\epsilon^2)$ in the patch coordinates, together they will contribute at $\mathcal{O}(\epsilon^3)$ and hence cannot contribute to the gravity equation at $\mathcal{O}(\epsilon^2)$. Similar arguments rule out coupling of the stress tensor with the first sub-leading order Taylor expansion coefficients of the shape and the velocity fields. Hence, we can study the contributions of the stress tensor and the curvatures of $ds_{asym}^2$ independent of the shape and velocity data. The metric corrections due to the data from the shape and velocity functions can be inferred from the corresponding results in \cite{Dandekar:2016fvw} and we will talk about it in a later section. 

\subsubsection{Tensor sector}
The metric correction in the tensor sector obeys the following differential equation
\begin{eqnarray}
\label{teneq}&&-\frac{n_s^2}{2S_0^2}\left((1-e^{-R})\frac{d^2}{dR^2}{\mathcal{T}}_{\mu\nu}+\frac{d}{dR}{\mathcal{T}}_{\mu\nu}\right)+\mathcal{S}^T_{\mu\nu}(R) =0,\\
\label{tensource}&&\text{where,}\quad  \mathcal{S}^T_{\mu\nu}(R)= \left( T^c_{\alpha\beta}P^\alpha_\mu P^\beta_\nu-T_{\alpha\beta}P^\alpha_\mu P^\beta_\nu\right)-e^{-R}R_{\alpha\gamma\beta\delta}O^\alpha O^\beta P^{\gamma}_\mu P^{\delta}_\nu.
\end{eqnarray}
The solution to the above differential equation can be inferred from the results presented in \cite{Dandekar:2016fvw} since we use the same set of boundary and regularity conditions. It is given by
\begin{eqnarray}\label{efftens}
	\mathcal{T}_{\mu \nu} &=& -\frac{2S_0^2}{n_s^2}\Bigg(\int_{R}^\infty\frac{dy}{e^y-1}\int_0^y e^x \mathcal{S}^T_{\mu\nu}(x) dx\Bigg).
\end{eqnarray}
The above integral is well-behaved in the limit $R\rightarrow\infty$. This is clear from the form of the outer integral if the corresponding integrand is regular in this limit. In this limit the integrand evaluates to
\begin{eqnarray}
	&&\lim_{y\rightarrow\infty}\frac{1}{e^y-1}\int_0^y e^x \mathcal{S}^T_{\mu\nu}(x) dx=\lim_{y\rightarrow \infty}\frac{\int_0^y e^x \mathcal{S}^T_{\mu\nu}(x) dx}{e^y}\quad (\because \lim_{y\rightarrow\infty}(e^y-1)=e^y)\nonumber\\
	&&=\lim_{x\rightarrow\infty}\mathcal{S}^T_{\mu\nu}(x)=0.\quad\\ &&(\because \text{applying L'Hopital rule and boundary conditions on stress tensor}. )\nonumber
\end{eqnarray}	
Since the integrand vanishes in this limit, the integral vanishes too and hence the tensor part of the metric correction is zero in the $R\rightarrow\infty$ limit. 

 Near $R\rightarrow0$ the integrand of the outer integral is well behaved provided the inner integral vanishes near $y\rightarrow0$ which is true by construction. Also, the integrand of the outer integral is regular in the $R\rightarrow\infty$ limit and hence, the metric correction is regular in the $R\rightarrow 0$ limit. Nothing special happens between these two limits and hence the metric correction in the tensor direction is regular everywhere in the membrane region. 

\subsubsection{Vector sector}
The differential equation that the vector metric correction $V_{Vi}$ satisfies is
\begin{eqnarray}
-\frac{n_s^2}{2S_0^2} \left(1-e^{-R}\right)\left(\frac{d^2}{dR^2}V_{Vi}+\frac{d}{dR}V_{Vi}\right)+\mathcal{S}_{Vi}(R)=0,
\end{eqnarray}
where,
\begin{eqnarray}
\mathcal{S}_{Vi}(R) = \left((1-e^{-R}){T}^c_{\mu\nu}(R)-{{T}}_{\mu\nu}\right)P^\mu_i u^\nu-e^{-R} R R_{\mu\nu\gamma\delta} n^{\mu} u^{\nu} O^{\gamma}{ P^\delta}_i.
\end{eqnarray}
Again using the general solution from \cite{Dandekar:2016fvw} and applying it to the particular source we get the following solution to the metric correction
\begin{eqnarray}\label{effvec1}
V_{Vi} &=& \frac{2S_0^2}{n_s^2}\Bigg(e^{-R}\int_0^R \frac{-e^x}{1-e^{-x}} \mathcal{S}_{Vi}(x)dx - \int_R^\infty \frac{1}{1-e^{-x}} \mathcal{S}_{Vi}(x) dx \nonumber\\
&&+ e^{-R} \int_0^\infty  \frac{1}{1-e^{-x}} \mathcal{S}_{Vi}(x) dx \Bigg)\nonumber\\
&=& \frac{2S_0^2}{n_s^2}\Bigg(-e^{-R}\int_0^R e^x\mathcal{S}_{Vi}(x)dx-(1-e^{-R})\int_R^\infty\frac{\mathcal{S}_{Vi}(x)}{1-e^{-x}}dx\Bigg).
\end{eqnarray}

From the boundary conditions on the membrane stress tensor near $R\rightarrow\infty$ we get
$$\lim_{R\rightarrow\infty}\mathcal{S}_{Vi}(R)=0.$$
Also, from the consistent solutions to the conservation of stress tensor in the membrane region we have $(u\cdot T)_\mu(R=0)=0$ and hence, 
$$\lim_{R\rightarrow 0}\mathcal{S}_{Vi}(R)=0.$$
Hence, in the limit of $R\rightarrow\infty$ the first integral above vanishes (by applying L'Hopital rule and using the boundary condition on $\mathcal{S}_{Vi}(R)$ near $R\rightarrow\infty$). Also, the second integral vanishes in the limit. Hence, $$\lim_{R\rightarrow \infty}V_{Vi}(R)=0.$$
In the limit of $R\rightarrow 0$ the first integral above vanishes. The second integral is finite in this limit since, the integrand vanishes near infinity and near $R=0$ the integrand is regular. Hence the second term also vanishes in this limit and we find that the metric correction satisfies the boundary condition 
$$\lim_{R\rightarrow 0}V_{Vi}(R)=0.$$
Moving on to the metric correction $V_{Zi}$, it satisfies the following second order differential equation
\begin{eqnarray}
-\frac{n_s^2}{2S_0^2} \left(\left(1-e^{-R}\right)\frac{d^2}{dR^2}V_{Zi}+\frac{d}{dR}V_{Zi}\right)+\mathcal{S}_{Zi}(R)=0,
\end{eqnarray}
where  $\mathcal{S}_{Zi}(R)=\frac{1}{1-n_s^2}\left(\left(\left(T^c-T\right).Z\right)_i+n_s e^{-R} R_{\mu\alpha\nu\beta}n^\mu u^\alpha O^\nu P^{\beta}_i\right)$. The solution to the above equation is 
 \begin{eqnarray}\label{vzisol}
\nonumber V_{Zi} &=&-\frac{2S_0^2}{{n_s}^2} \Big(\int_R^\infty \frac{dy}{e^y-1} \int_0^y e^x  \mathcal{S}_{Zi}(x) dx\Bigg).
\end{eqnarray}
We see that the expression of the solution to $V_{Zi}$ is similar to the solution for the metric correction in the tensor sector and hence from analogy with that solution we can conclude that $V_{Zi}$ is regular everywhere in the membrane region and vanishes in the $R\rightarrow \infty$ limit. 

\subsubsection{Scalar Sector}
In the scalar sector the metric corrections $S_{VZ}(R)$ is obtained by solving the second order differential equation given by 
\begin{eqnarray}
-\frac{n_s^2}{2S_0^2} \left(1-e^{-R}\right)\left(\frac{d^2}{dR^2}S_{Vz}+\frac{d}{dR}S_{Vz}\right)+\mathcal{S}_{Vz}(R)=0,
\end{eqnarray}
where we divide the source in two parts as $\mathcal{S}_{Vz}(R) = \mathcal{S}^1_{Vz}(R)+\mathcal{S}^2_{Vz}(R)$ with
\begin{eqnarray}
&& \mathcal{S}^1_{Vz}(R) = \Bigg(
- \frac{n_s(2-3n_s^2)}{6(1-n_s^2)^2}R_{\alpha\beta\gamma\delta}n^\alpha u^\beta n^\gamma u^\delta\nonumber\\
 && +\frac{n_s}{3(2-n_s^2)}\left(\frac{1}{1-n_s^2}(u.T^c.u+n.T^c.n)+Tr[T^c.P]\right)\Bigg)e^{-2R}(e^R-1),
\end{eqnarray}
and
\begin{eqnarray}
 \mathcal{S}^2_{Vz}(R) = \frac{1}{1-n_s^2}\left((1-e^{-R})(u.T^c.Z) - (u.T.Z) \right)+\frac{n_s}{1-n_s^2}R e^{-R}R_{\alpha\beta\gamma\delta}n^\alpha u^\beta n^\gamma u^\delta,\nonumber
\end{eqnarray}
where, $Tr[T^c.P]$ is the trace of the stress tensor components orthogonal to $n$, $u$ and $Z$ vectors in the effective spacetime directions. We have split up the source term into two parts to help with our analysis of 'geometrisation' of the metric corrections later. The solution for $S_{Vz}$ is given by
\begin{eqnarray}
S_{Vz} &=&- \frac{2S_0^2}{n_s^2}\Bigg(e^{-R}\int_0^R e^x \mathcal{S}_{Vz}(x)dx +(1-e^{-R})\int_R^\infty \frac{\mathcal{S}_{Vz}(x)}{1-e^{-x}} dx \Bigg).
\end{eqnarray}
The source $\mathcal{S}_{Vz}(R)$ vanishes at both the limit $R\rightarrow 0$ and $R\rightarrow \infty$. The form of the solution $S_{Vz}$ is same as \eqref{effvec1}, thus it can be shown that $S_{Vz}$ will also vanish at both asymptote and at $R\rightarrow 0$.

The metric correction $S_{Tr}$ is obtained by solving the following differential equation
\begin{eqnarray}
-\frac{n_s^2}{S_0^2}\left((e^{-R}-1)\frac{d^2}{dR^2}S_{Tr}+\frac{d}{dR}S_{Tr}\right)+\mathcal{S}^{Tr}_T(R)=0,
\end{eqnarray}
where 
\begin{eqnarray}
\mathcal{S}_T^{Tr}(R) &=& 
\Bigg(Tr(P.T^c)-Tr(P.T)+2(\bar{T}-\bar{T}^c)-e^{-R} O.T^c.O\nonumber\\&&+\frac{n_s^2}{1-n_s^2}e^{-R}R_{\alpha\mu\beta\nu}n^\alpha u^\mu n^\beta u^\nu\Bigg).
\end{eqnarray}
The solution of this equation is given by
\begin{equation}
S_{Tr} =- \frac{S_0^2}{n_s^2} \left( \int_R^\infty \frac{dy}{e^y-1} \int_0^y e^x \mathcal{S}^{Tr}_T(x)  dx\right).
\end{equation}
By the similarity of the form of the solution of $S_{TR}$ with \eqref{efftens}, we know that the metric correction satisfies the right boundary and regularity conditions. 

The metric correction $S_{ZZ}$ satisfies the differential equation 
 \begin{eqnarray}
-\frac{n_s^2}{2S_0^2} \left(\left(1-e^{-R}\right)\frac{d^2}{dR^2}S_{ZZ}+\frac{d}{dR}S_{ZZ}\right)+\mathcal{S}_{ZZ}(R)=0,
\end{eqnarray}
where
\begin{eqnarray}
\mathcal{S}_{ZZ} (R)&=& \frac{1}{(1-n_s^2)^2}\left(Z.(T^c-T).Z - n_s^2e^{-R} R_{\mu\alpha\nu\beta} n^\mu u^\alpha n^\nu u^\beta \right)\nonumber\\&&-\frac{(\bar{T}^c-\bar{T})}{(1-n_s^2)},
\end{eqnarray}
and
\begin{eqnarray}\label{vzzsol}
 S_{ZZ} &=& -\frac{2S_0^2}{{n_s}^2} \left(\int_R^\infty \frac{dy}{e^y-1} \int_0^y e^x \mathcal{S}_{ZZ}(x)dx \right).
\end{eqnarray}
From the property of the stress tensor at the asymptote, the source $\mathcal{S}_{ZZ}(R) \rightarrow 0$ at the asymptote. As the form of the solution is similar to the \eqref{efftens}, and also the form of the source mimics the behaviour of \eqref{tensource}, $S_{ZZ}\rightarrow 0$ as $R\rightarrow \infty$ and it is regular as   $R\rightarrow 0$.

Next we find that the dilaton correction $\delta\phi^{(2)}$ satisfies the following differential equation
 \begin{equation}
 \frac{d^2\delta\phi^{(2)}}{dR^2}+(1-n_s^2)\frac{d^2S_{ZZ}}{dR^2}+2 \frac{d^2S_{Tr}}{dR^2}+2\frac{S_0^2}{n_s^2}\left( O.T.O(R)-O.T^c.O\right)=0.
 \end{equation}
The solution to the above equation is
\begin{eqnarray}
\delta\phi^{(2)} 
&=&2\int_R^\infty \left(\frac{S_0^2}{n_s^2}x \left(O.T^c.O- O.T.O(x)\right)\right)dx-2R\int_R^\infty \left(\frac{S_0^2}{n_s^2} \left(O.T^c.O- O.T.O(x)\right)\right)dx\nonumber\\&&-2S_{Tr}(R)-(1-n_s^2)S_{ZZ}(R).
\end{eqnarray}
Both the integrands above vanish in the $R\rightarrow\infty$ limit by the asymptotic boundary condition on the stress tensors and as a result  the two integrals vanish at $R\rightarrow\infty$. In addition since both of $S_{ZZ}(R)$ and $S_{Tr}(R)$ are also zero in this limit, we have $\delta\phi^2\rightarrow 0$ as $R\rightarrow\infty$. In the $R\rightarrow 0 $ limit both the integrands are regular. Also, since the integrands vanish near $R\rightarrow\infty$, both the integrals are finite in the $R\rightarrow 0 $ limit. Hence, with the information that both $S_{ZZ}$ and $S_{Tr}$ are regular near $R\rightarrow 0$, we find that $\delta\phi^2$ is also regular near $R\rightarrow 0 $.

Once, we have solved for the other metric corrections in the scalar sector we find that $S_{VV}$ satisfies the following first order differential equation.
\begin{eqnarray}
-\frac{n_s^2}{2S_0^2}\left(\frac{d}{dR}S_{VV}+S_{VV}\right)+\mathcal{S}_{vv}(R)=0,
\end{eqnarray}
where
\begin{eqnarray}
&&\mathcal{S}_{vv}(R) = \frac{n_s (n_s^2-1)}{2S_0^2} \left(\frac{d}{dR} S_{VZ}(R)+2S_{VZ}(R)\right)\nonumber\\&&+(2-e^{-R})\frac{n_s^2}{4S_0^2}\frac{d}{dR}\left(2S_{Tr}(R)-(n_s^2-1)S_{ZZ}(R)+\delta\phi^2(R)\right) -\frac{(n_s^2-1)^2}{2S_0^2}S_{ZZ}(R) \nonumber\\&& -(1-e^{-R}) O.T.O (R) - O.T.u (R) -\frac{3+R e^{-R}}{3} u.T^c.n+\left(1-\frac{R e^{-R}}{2}\right)n.T^c.n\nonumber\\
&&+\frac{e^{-R}(R-2)}{6}u.T^c.u +\frac{e^{-R}(2R-1)}{3} \bar{T}^c +\frac{e^{-R}(4-18R+3R^2)}{12} R_{\alpha\mu\beta\nu}n^\alpha u^\mu n^\beta n^\nu \nonumber\\&&
-\frac{(R+n_s^2)e^{-R}}{6(1-n_s^2)}R_{\alpha\mu\beta\nu}n^\alpha u^\mu n^\beta n^\nu. 
\end{eqnarray}
The solution to the above equation is given by
\begin{eqnarray}
S_{VV} &=& \frac{2S_0^2 e^{-R}}{n_s^2}\int_0^R e^{x} \mathcal{S}_{vv}(x)dx.
\end{eqnarray}
Near $R\rightarrow 0$ the above integral vanishes provided the integrand is regular in this limit.~Since the metric corrections are regular near $R\rightarrow 0$ and since the stress tensor components are also regular everywhere in the membrane region, we find that $\mathcal{S}_{vv}(R)$ is indeed regular in this limit and $S_{VV}(R=0)$ is zero and hence the boundary condition on $S_{VV}(R)$ is satisfied. In the limit $R\rightarrow\infty$, the integrand vanishes ( due to the asymptotic  boundary conditions on the metric corrections and the stress tensors). The integral is finite and hence $S_{VV}(R)$ goes to zero in the limit $R\rightarrow\infty$.

\subsection{Geometrisation}
We have briefly mentioned earlier that the solutions to the effective metric and dilaton field presented so far are invariant under change of number of dimensions included in $p$. This is because a solution which has $SO(D-p-2)$ isometry necessarily also has an $SO(D-q-2)$ isometry when $p<q$. In \cite{Bhattacharyya:2015fdk} it was shown that this property is manifest in the black hole solutions because we can recast the effective metric and dilaton field together into a metric correction of the full spacetime which is independent of $p$. This method was referred to as ``geometrisation" in \cite{Bhattacharyya:2015fdk}. The general structure of the metric corrections in full spacetime (with our choice of gauge) in a manifestly $p$ independent manner is given by \cite{Bhattacharyya:2015fdk, Dandekar:2016fvw},
\begin{eqnarray}
	&&H_{MN}=H^{(S)} O_M O_N+\frac{1}{D}H^{(Tr)} p_{MN}+H^{(V)}_{(M}O_{N)}+H^{(T)}_{MN},\nonumber\\
	\text{where,}\quad && p_{MN}=g^{asym}_{MN}+u_M u_N+n_M n_N,\nonumber\\
	\text{and,}\quad && H^{(V)}_M u^M= H^{(V)}_M n^M=0, H^{(T)}_{MN}u^M=H^{(T)}_{MN}n^M=H^{(T)}_{MN}p^{MN}=0.\nonumber\\
\end{eqnarray}
The assumption that we have made in writing the above metric correction is that near any arbitrary point in the membrane region, the only two distinct directions are along $d\psi$ and along the generator of the even horizon $u^M\partial_M$.~The only other assumption is that the dynamics are confined along a finite number of directions.

In \cite{Bhattacharyya:2015fdk} a map was obtained between the metric and dilaton corrections in the effective space-time and the metric corrections in the global coordinates under geometrisation. This map gets slightly modified due to the presence of curvature of the asymptotic space-time and the matter stress tensor parametrising our solutions. We present the modified map here in Table\eqref{tab:table2}. In Table\eqref{tab:table2} $P_{MN}$ denotes the projector orthogonal to $n, u,Z$ as well as the $D-p-3$ isometry directions.

\begin{table}[h!]
  \begin{center}
 \caption{Geometrisation relation}
    \label{tab:table2}
    \begin{tabular}{|l|c|r} 
  \hline
   \textbf{In Effective coordinate} & \textbf{Geometrised value}\\\hline
    $S_{VV}+ \frac{S_0^2}{n_s^2}e^{-R} R^2 (2 R_{\mu\alpha\nu\beta}n^\mu u^\alpha n^\nu u^\beta-\frac{2}{3}u.T^c.n)$ & $H_s$ \\\hline
    $S_{VZ}-\frac{(2-3n_s^2)S_0^2e^{-R}R}{3n_s(1-n_s^2)^2}R_{\mu\alpha\nu\beta}n^\mu u^\alpha n^\nu u^\beta$ & $H^V_M Z^M$ 
    \\ $+\frac{2S_0^2e^{-R}R}{3n_s(1-n_s^2)}(u.T^c.u+\bar{T^c})$ & 
    \\\hline
    $S_{ZZ}$ & $H^T_{AB} Z^A Z^B$\\\hline
     $p S_{tr}$ & $H^T_{AB} P^{AB}$\\\hline
    $V_{Vi}$ & $H^V_M P^M_i$\\\hline
    $V_{Zi}$ & $H^T_{AB}  Z^A P^B_i$\\\hline
    $\mathcal{T}_{MN} $ & $P_{M}^A P_{N}^B H^T_{AB}-\frac{{P_{MN}}}{p}P^{AB}H^T_{AB}$\\\hline
     $ \delta\phi^{(2)}+ p S_{tr}+(1-n_s^2)S_{ZZ}$ & $H^{Tr}$\\\hline
       \end{tabular}
  \end{center}
\end{table}
Using this map and the expression of the metric and dilaton corrections in the effective spacetime directions obtained in previous sections, we arrive at the following expressions of the metric corrections in the global coordinates (classified according to the tensor structure of the hyperplane orthogonal to $u^M$ and $d\psi$). We find that even though the map between effective spacetime metric and dilaton and the geometrised metric changes in presence of external forcing due to the matter stress tensor, the final geometric metric can be consistently written in the form mentioned above. i.e even the expressions for the metric corrections are independent of any reference to the $Z^\mu$ direction. 
\subsubsection{Geometrisation of the leading order ansatz}
Before going to geometrise the sub-leading corrections that we obtain in the presence of stress tensor, here we write down the first order corrected metric in geometrised fashion as follows
\begin{eqnarray}
&&g^{(1)}_{MN} = g^{asym}_{MN}+\frac{O_{M}O_{N}}{\psi^{D-3}} \nonumber\\&&+\frac{1}{D-3} \Big( \frac{D-3}{\mathcal{K}}R e^{-R}\Big(R\Big(-\frac{\mathcal{K}}{D-3}-\frac{u.\nabla \mathcal{K} }{\mathcal{K}}+\frac{u.\mathcal{K}.u}{2}\Big)
	+\Big(\frac{\mathcal{K}}{D-3}+u.\mathcal{K} .u\Big)\Big)O_{M} O_{N} \nonumber\\&&+\frac{D-3}{\mathcal{K}}R e^{-R}\big(u^{A} \mathcal{K}_{AC}-u^{A}\nabla_{A} u_{C}\big)P^C_{(M}O_{N)}\Big).\nonumber\\
\end{eqnarray}
In the above expression, $R$ should be thought of as being equivalent to $\frac{\psi-1}{D}$ and $e^{-R}$ should be thought of as being equivalent to $\psi^{-D}$. $\mathcal{K}_{AB}$ is the extrinsic curvature of the constant $\psi$ slices where the metric is evaluated. 
\subsubsection{Tensor Sector}
The solution for the geometric form of the tensor part of the metric corrections is given by
\begin{eqnarray}\label{mettens}
\nonumber H^T_{MN} &=&- \frac{2(D-3)^2}{{\mathcal{K}}^2} \left(\int_R^\infty \frac{dy}{e^y-1}\int_0^y e^x \mathcal{S}^T_{MN}(x) dx\right),
\end{eqnarray}
where $\mathcal{S}^T_{MN}(R)$ is given  by
\begin{eqnarray}
\mathcal{S}^T_{MN}(R) &=& T^c_{AB} p^A_M p^B_N-T_{AB}(R) p^A_M p^B_N-e^{-R} R_{ACBD}O^AO^B p^C_M p^D_N \nonumber\\&&+\mathcal{\tilde{S}}^T_{MN}(R),\nonumber\\
\text{where,}\quad && R=D\left(\psi-1\right).
\end{eqnarray}
Here, $T^c_{AB}$ and $T_{AB}(R)$ are the components of the asymptotic stress tensor and the stress tensor in the membrane region. $p_{MN}$ is the projector orthogonal to $u_M$ and $n_M$. $\mathcal{\tilde{S}}^T_{MN}(R)$ is the part of the source contributing to the differential equation for $H^{(T)}_{MN}$ due to the membrane shape and velocity data. This part of the source can be inferred from the sources mentioned in \cite{Dandekar:2016fvw} at the second sub-leading order in $1/D$ by interpreting all covariant derivatives and dot products now being taken w.r.t the asymptotic metric $g^{asym}_{MN}$ here. The fact that the solution mentioned above is well behaved everywhere in the membrane region can be inferred from the analysis of the corresponding integral solutions in the effective spacetime in the earlier section.

\subsubsection{Vector Sector}
The solution of the metric corrections in the vector direction is given by
\begin{eqnarray}\label{metvec}
\nonumber H^V_{M}(R) &=& -\frac{2(D-3)^2}{{\mathcal{K}}^2}\Bigg(e^{-R}\int_0^R e^x \mathcal{S}^V_{M}(x)dx +(1-e^{-R})\int_R^\infty \frac{\mathcal{S}^V_{M}(x) }{1-e^{-x}} dx \Bigg),
\end{eqnarray}
where 
\begin{eqnarray}
\mathcal{S}^V_{M}(R) &=& e^{-R}\left((e^R-1){T}^c_{AB}-e^R{{T}}_{AB}\right)p^A_M u^B-Re^{-R} R_{ABCN}n^{A}u^{B}O^{C}p^N_M\nonumber\\
&&+\mathcal{\tilde{S}}^V_{M}(R).
\end{eqnarray}
Once again, $\tilde{\mathcal{S}}^{(V)}_M$ denotes the part of the source coming from the membrane shape and velocity data in the differential equation for $H^{(V)}_M$. 
\subsubsection{Scalar Sector}
The scalar part of the metric has two different components. The solution for the trace component along $p^{MN}$ is given by
\begin{eqnarray}
H^{Tr} &=& 2\frac{(D-3)^2}{\mathcal{K}^2}\left(\int_R^\infty x \mathcal{S}^{Tr}(x)dx -R\int_R^\infty \mathcal{S}^{Tr}(x)dx\right),
\end{eqnarray}
where $\mathcal{S}^{Tr}(R)=\left(O.T^c.O- O.T.O(R)\right)+\tilde{\mathcal{S}}^{Tr}(R)$.
Here also the limit of the integral is fixed by requiring the condition that metric correction should vanish at $R\rightarrow \infty$.
The solution for the second component of the scalar metric correction is given by
\begin{equation}\label{scalmet}
H^s(R) =2e^{-R} \frac{(D-3)^2}{\mathcal{K}^2} \int_0^R e^{x} \mathcal{S}^S(x) dx,
\end{equation}
where,
\begin{eqnarray}
&&\mathcal{S}^S(R) =  -\frac{\mathcal{K}}{2(D-3)}\frac{d}{dR}(\nabla^M H^V_M) -\frac{\mathcal{K}}{(D-3)}\nabla^M H^V_M+\frac{\mathcal{K}^2}{4(D-3)^2}(2-e^{-R})\frac{d}{dR}H^{Tr}\nonumber\\
&&-\frac{1}{2}\nabla^M \nabla^N H^T_{MN}-(1-e^{-R})O.T.O(R)-O.T.u(R)+\frac{R^2}{4}e^{-R} R_{\alpha\mu\beta\nu}n^\alpha u^\mu n^\beta u^\nu
\nonumber\\
&& -(1+R e^{-R}) u.T^c.n+\frac{1}{2}Re^{-R} u.T^c.u+\left(1-\frac{Re^{-R}}{2}\right) n.T^c.n+e^{-R}R \bar{T}^c+\tilde{\mathcal{S}}(R).\nonumber\\
\end{eqnarray}
The limits of integration used above are to ensure that the metric correction $H^{(S)}$ satisfy the boundary condition that it vanishes at $R=0$ and also that it has non-zero support only in the membrane region.

\subsection{The sub-leading order corrections to the membrane equations }
The constraints on the membrane variables necessary for the constraint Einstein equations to be satisfied gets corrected at this order. This in turn gives rise to sub-leading order in $1/D$ corrections to the membrane equations. Once we use the fact that conservation of the stress tensor in the membrane region requires that $$(u\cdot T)(R=0)=0,$$ we find that the dependence of the constraint equation on the matter stress tensor in the membrane region vanishes. What we are left with is a set of constraint on the membrane variables for the membrane propagating in the asymptotic spacetime parametrised by the asymptotic stress tensor and the curvature tensor of the asymptotic space-time. These constraints when written in terms of the variables of the membrane world-volume gives rise to the following sub-leading order corrected membrane equations

\begin{equation}\label{eq:constraint223}
	\begin{split}
		&\left[\frac{\hat{\nabla}^2 u_A}{\cal{K}}-\frac{\hat{\nabla}_A{\cal{K}}}{\cal{K}}+u^B {\cal{K}}_{BA}-u\cdot\hat{\nabla} u_A\right]{\cal P}^A_C+ \Bigg[-\frac{u^B {\cal{K}}_{B D} {\cal{K}}^D_A}{\cal{K}}+\frac{\hat{\nabla}^2\hat{\nabla}^2 u_A}{{\cal{K}}^3}-\frac{(\hat{\nabla}_A{\cal{K}})(u\cdot\hat{\nabla}{\cal{K}})}{{\cal{K}}^3}\\
		&-\frac{(\hat{\nabla}_B{\cal{K}})(\hat{\nabla}^B u_A)}{{\cal{K}}^2}-\frac{2{\cal{K}}^{D B}\hat{\nabla}_D\hat{\nabla}_B u_A}{K^2} -\frac{\hat{\nabla}_A\hat{\nabla}^2{\cal{K}}}{{\cal{K}}^3}+\frac{\hat{\nabla}_A({\cal{K}}_{BD} {\cal{K}}^{BD} {\cal{K}})}{K^3}+3\frac{(u\cdot {\cal{K}}\cdot u)(u\cdot\hat{\nabla} u_A)}{{\cal{K}}}\\
		&-3\frac{(u\cdot {\cal{K}}\cdot u)(u^B {\cal{K}}_{BA})}{{\cal{K}}}
		-6\frac{(u\cdot\hat{\nabla}{\cal{K}})(u\cdot\hat{\nabla} u_A)}{{\cal{K}}^2}+6\frac{(u\cdot\hat{\nabla}{\cal{K}})(u^B {\cal{K}}_{BA})}{{\cal{K}}^2}+3\frac{u\cdot\hat{\nabla} u_A}{D-3}\\
		&-3\frac{u^B {\cal{K}}_{BA}}{D-3}\Bigg]{\cal P}^A_C +\frac{T^c_{AB} u^B}{\cal{K}}{\cal P}^A_C = {\cal{O}}\left(\frac{1}{D}\right)^2,
		\\
		\\
		&\hat{\nabla}\cdot u = \frac{1}{2{\mathcal K}}\left( \hat{\nabla}_{(A}u_{B)}\hat{\nabla}_{(C}u_{D)}{\cal P}^{BC}{\cal P}^{AD} \right)+ {\cal{O}}\left(\frac{1}{D}\right)^2. 
	\end{split}
\end{equation} 

Here again, all covariant derivatives are evaluated w.r.t the induced metric on the membrane propagating in the asymptotic spacetime of the black hole. In terms of variables of the membrane world-volume, the only change in the above equation w.r.t the sub-leading order membrane equations evaluated in \cite{Dandekar:2016fvw} is the presence of the term proportional to $T^c\cdot u$ in the vector membrane equations. 

\section{Conclusions}
In this paper, we have derived large $D$ dynamical black hole solutions which asymptote to  arbitrary spacetimes which are solutions of Einstein equations with very weak matter stress tensors. The weakness is parametrised by the fact that the Riemann curvature tensor of the asymptote is $\mathcal{O}(D^0)$ to the leading order in large $D$. The metric is obtained in terms of the variables of a dual co-dimension one membrane propagating in a spacetime equivalent to the asymptotic spacetime of the black hole. In addition, the black hole solution also depends on the curvatures of the asymptotic spacetime, the matter stress tensor in the `membrane region' and the asymptotic value of the matter stress tensor outside the membrane region. 

We have obtained the metric of the black hole up to the second sub-leading order in $1/D$. The metric solutions are accompanied by a set of equations constraining the dynamics of the dual membrane.~We call these the membrane equations and we have obtained the expression of the membrane equations up to the first sub-leading order in large $D$. Quite conveniently we find that the membrane equation depends only on the data of the membrane propagating in the asymptotic spacetime. More precisely this equation is independent of the form of the stress tensor in the membrane region. The crucial input which made this happen is the conservation of the stress tensor which is a requirement for the stress tensor to be consistently coupled with gravity. 

The solution we have obtained are very general in the sense that we have only demanded that the matter stress tensor is regular, conserved and that it is able to source only Riemann curvatures of strength $\mathcal{O}(D^0)$ in the asymptote. Due to the weakness of the stress tensor, the evolution of the stress tensor in a time like direction is not relevant at this order in $1/D$ as this information enters nowhere in either the solution of the metric or in the expression of the membrane equations.  Hence, the fact that we do not have membrane equations corresponding to the matter sector \footnote{ E.g. see \cite{Bhattacharyya:2015fdk} for membrane equations determining the evolution of charge density of the membrane for a stronger stress tensor in the membrane region which though has non-zero support only in the membrane region.} is not a problem for us since we are effectively agnostic to the dynamics of the matter sector. Hence, for any matter stress tensor with the above mentioned properties, the metric solutions obtained by us are a complete description of the dynamics of the black hole. 

Owing to the weakness of the matter stress tensor we are not able to probe any possible non-trivial constraints on the stress tensor for the black hole to satisfy the second law of thermodynamics. It will be interesting to probe the effect of a stronger stress tensor which sources curvatures in the asymptotic spacetime of the order of $\mathcal{O}(D)$. This stress tensor will see more of the finer structure of the black hole metric and may have additional non-trivial constraints on it from the consistency with second law (beyond the null-energy condition). 

Probing spacetimes with curvatures of the order of $\mathcal{O}(D)$ requires that the derivatives on the metric be $\mathcal{O}(\sqrt{D})$.  It is expected that the membranes propagating in these spacetimes will have extrinsic curvatures and derivatives on the velocity field which are the order of $\mathcal{O}(\sqrt{D})$.  This will be a new length scale to probe in the large $D$ regime from the membrane perspective. This regime has been probed in the effective mass-momentum picture of  \cite{Emparan:2015gva,Tanabe:2015hda,Emparan:2016sjk,Tanabe:2016pjr} and will help us to further explore the equivalence (or lack of it) between the membrane and the mass-momentum formalism to study effective non-gravitational systems dual to large $D$ black holes. 

The counterpart of our analysis in a spacetime with negative cosmological constants is equivalent to introducing deformations in the boundary metric of the spacetime. The effective equations in presence of boundary deformations in the mass momentum formalism has been computed in \cite{Andrade:2018zeb} and the fluid like behaviour of the corresponding non-gravitational system has been studied in \cite{Andrade:2019rpn}. To obtain equivalent results in the membrane formalism we will need to do the analysis in presence of strong curvature asymptotes (mentioned in the previous paragraph) in presence of cosmological constant. We leave these directions of explorations to future projects.

\section{Acknowledgement}
 The work of TM is supported by the Simons Foundation Grant Award ID 509116 and by the South African Research Chairs initiative of the Department of Science and Technology and the National Research Foundation. The work of AS is supported by the Ambizione grant no.  $PZ00P2\_174225/1$   of the Swiss National Science Foundation (SNSF) and partially by the NCCR grant no. $51NF40-141869$  “The Mathematics of Physics” (SwissMap). TM would also like to thank the hospitality of University of Geneva during a visit where this project was initiated. 
\appendix
\section*{Appendix}
\section{Scaling properties of various matter stress tensors}\label{Emax}
In this appendix we study the scaling properties of the fundamental fields of different matter contents which give rise to the scaling properties of the stress tensor required by us. 
\subsubsection*{The Einstein-Maxwell Stress tensor}
Due to the isometry of the configurations that we are studying the vector potential, $A_M$ has dependence only along the effective $p+3$ dimensional spacetime and also has its non-zero components only along these directions.~Hence the field strength $F_{MN}$ to have components only along the $x^\mu$ directions. 
i,e, $$F_{i\mu}=F_{ij}=0.$$
The expression for the electromagnetic stress tensor is given by
\begin{equation}
T_{AB}=F_{AC}F_B^C-\frac{1}{4}g_{AB}F_{CD}F^{CD}.
\end{equation}
Hence, the component of the stress tensor along the large sphere directions are given by
\begin{eqnarray}
T_{ij}&=&-\frac{1}{4}e^{\phi}\Omega_{ij}F_{\mu\nu}F^{\mu\nu},\nonumber\\
\implies \bar{T}&=&-\frac{1}{4}F_{\mu\nu}F^{\mu\nu}.
\end{eqnarray}
 The component of the stress tensor along the effective spacetime directions are given by
\begin{eqnarray}
T_{\mu\nu}=g^{\gamma\delta}F_{\mu\gamma}F_{\nu\delta}-\frac{1}{4}g_{\mu\nu}F_{\alpha\beta}F^{\alpha\beta}.
\end{eqnarray}
Under the coordinate transform from the global $x^\mu$ coordinates to the patch $y^a$ coordinates the fields transform as
\begin{eqnarray}
&&g_{ab}=\frac{\partial x^\mu}{\partial y^a}\frac{\partial x^\nu}{\partial y^b}g_{\mu\nu}=\frac{1}{(D-3)^2}\alpha^\mu_a\alpha^\nu_bg_{\mu\nu},\nonumber\\
&&\partial_a\phi=\frac{1}{D-3}\alpha^\mu_a\partial_\mu\phi,\nonumber\\
&&A_a=\frac{1}{D-3}\alpha^\mu_a A_\mu.
\end{eqnarray}
The rescaling necessary to capture the type of solutions that we mentioned about earlier is given by
\begin{eqnarray}
 &&G_{ab}=(D-3)^2g_{ab}=\alpha^\mu_a\alpha^\nu_bg_{\mu\nu},\nonumber\\
&&G^{ab}=\frac{1}{(D-3)^2}g^{ab}=\alpha^a_\mu\alpha^b_\nu g^{\mu\nu},\nonumber\\
&&\chi_a=(D-3)\partial_a\phi=\alpha^\mu_a\partial_\mu\phi,\nonumber\\
&&A^{(G)}_a=(D-3) A^{(g)}_a.
\end{eqnarray}
 Under the subsequent action of the scaling of the fields, the relevant quantities constituting the stress tensor for the Maxwell fields transform as 

\begin{eqnarray}
	&& g^{cd}F^{(g)}_{ac}F^{(g)}_{bd}\rightarrow \frac{F^{(G)}_{ac}}{D-3}\frac{F^{(G)}_{bd}}{D-3}G^{cd}(D-3)^2=F^{(G)}_{ac}F^{(G)}_{bd}G^{cd},\nonumber\\
	&&\frac{1}{4}F^{(g)}_{ab}F^{(g)ab}\rightarrow\frac{(D-3)^2}{4}F^{(G)}_{ab}F^{(G)ab}.
\end{eqnarray}
Under these transformations, the components of the stress tensor transform as
\begin{eqnarray}
&&T^{(g)}_{ab}\rightarrow T^{(G)}_{ab},\nonumber\\
&&\bar{T}^{(g)}\rightarrow (D-3)^2\bar{T}^{(G)}.
\end{eqnarray}
Hence, the scaling of the vector potential $A_M^{(G)}$ assumed above gives rise to the scaling properties of the stress tensor that we require. 
\subsection*{Ideal Fluid}
The stress tensor for ideal fluid at zeroth order in derivatives is given by 
$$T_{\mu\nu} = (E+P) u_\mu u_\nu + P g_{\mu\nu},$$
where $E$ is the energy density and $P$ is the pressure, $v^\mu$ is the velocity field of the fluid. To satisfy the above mentioned scaling properties of the stress tensor, $v^\mu$ , $E$ and $P$ must scale as
\begin{eqnarray}
&&P^{(g)}\rightarrow D^2 P^{(G)},\\
&&E^{(g)}\rightarrow D^2 E^{(G)},\\
&&u_\mu^{(g)}\rightarrow \frac{1}{D} u^{(G)}_\mu.
\end{eqnarray}
\subsection*{Cosmological constant}
 The effective stress tensor for a cosmological constant which contributes to an intrinsic curvature of inverse length scale of $\mathcal{O}(D^0)$ is given by \cite{Bhattacharyya:2017hpj,Bhattacharyya:2018szu}
\begin{equation}
T^{cosm}_{\mu\nu}=-\frac{(D-1)(D-2)}{l^2}g_{\mu\nu}.
\end{equation}
Hence it shares the scaling properties of the metric which written explicitly is given by
\begin{eqnarray}
&& T^{(g)cosm}_{ab}\rightarrow \epsilon^2T^{(G)cosm}_{ab},\nonumber\\
&&\bar{T}^{(g)cosm}\rightarrow \bar{T}^{(G)cosm}.
\end{eqnarray}
Thus, the scaling property required by us is not satisfied by the stress tensor of the cosmological constant. \footnote{The scaling property of the stress tensor can be matched with the one required by us by scaling the variable $`'l`$ but then we loose the property that the asymptote has a non-trivial curvature of the order of $\mathcal{O}(D^0)$. }

\section{How $\mathcal{O}(D^0$) stress tensor generates $\mathcal{O}(D^0$) internal curvatures}\label{weakstress}
In this appendix we show that with the symmetry properties of the solution that we are considering, $\mathcal{O}(D^0)$ stress tensor give rise to internal curvatures which are $\mathcal{O}(D^0)$. More generally the strength of the stress tensor is the same as the strength of all the curvature components. The metrics that we work with in this paper can be written in the effective spacetime language in a manner which manifests the $SO(D-p-2)$ isometry of the spacetime. It is given by 
\begin{equation}
	ds^2=\mathcal{G}_{MN}dx^M dx^N=g_{\mu\nu}(x) dx^\mu dx^\nu +e^\phi(x) d\Omega_{D-p-3}^2.
\end{equation}
Let ${M,N,\ldots}$ denote the coordinate directions along the full spacetime, ${i,j,\ldots}$ denote the coordinates along the $d\Omega_{D-p-3}^2$ direction and ${\mu,\nu,\ldots}$ denote the coordinate directions along the effective spacetime. With this convention, the non-zero components of the Riemann curvatures of $\mathcal{G}_{MN}$ is given by
\begin{eqnarray}
	&&R^\alpha_{\beta\gamma\delta}=\bar{R}^\alpha_{\beta\gamma\delta},\nonumber\\
	&&R^\alpha_{i\beta j}=\left(-\frac{1}{4}\bar{\nabla}_\beta \phi\bar{\nabla}^\alpha\phi-\frac{1}{2}\bar{\nabla}_\beta \bar{\nabla}^\alpha\phi\right)\Omega_{ij}e^\phi,\nonumber\\
	&&R^i_{jkl}=\left(1-\frac{1}{4}\partial_\mu \phi \partial^\mu \phi e^\phi\right)\left(\delta^i_k\Omega_{jl}-\delta^i_l \Omega_{jk}\right),
\end{eqnarray}
where, the quantities with overhead bars are to be thought of as being evaluated w.r.t. the effective metric $g_{\mu\nu}$. The corresponding expressions of the curvatures in a patch like coordinate for the full spacetime will also have the same structure as above. In the patch-like coordinates we trade the variables $\partial_a\phi$ with $\chi_a$.~With a particular choice of patch coordinates, the $ds_{asym}^2$ part of the ansatz metric around any point $x_0$ in the membrane region can be written as follows
\begin{equation}
	g^{asym}_{MN}=\eta_{MN}+\frac{\epsilon^2}{3}R_{MANB} y^A y^B.
\end{equation}
This choice of patch coordinate is the Riemann normal coordinate of the full spacetime. 
For $g^{asym}_{MN}$ to maintain the effective spacetime structure mentioned above we need to have $$g_{ij}\propto \Omega_{ij}\implies R_{ijkl}=0.$$
This imposes the following constraints on the dilaton field in the non-patch (global) coordinates
\begin{equation}
	\partial_\mu \phi \partial^\mu \phi=4 e^{-\phi}.
\end{equation}
Similarly, the effective spacetime structure of the full metric requires that $$g_{i\mu}=0\implies R_{\alpha j i \beta}=0.$$ This imposes a further constraint on the dilaton field, namely
\begin{equation}
	\bar{\nabla}_\beta\phi\bar{\nabla}_\alpha\phi+2\bar{\nabla}_\beta\bar{\nabla}_\alpha\phi=0.
\end{equation}
So, in a local coordinates in which we are able to express $ds_{asym}^2$ in terms of Riemann normal coordinates, the only non-zero components of the Riemann tensor at $x_0$ are $R_{\alpha\beta\gamma\delta}=\bar{R}_{\alpha\beta\gamma\delta}$. 

In a spacetime with large isometry contractions of tensors usually adds an order in $D$ to the naive order in $D$ of the tensor by virtue of contractions with the metric of the isometry sphere directions. But for tensors which do not have non-zero components along the isometry sphere directions, we do not get this increase in order of $D$ due to contraction. We have shown above that in a particular coordinate system the components of the Riemann tensor are zero along the sphere directions. Since, we only allow diffeomorphisms which do not mix the effective spacetime directions and the large sphere directions (in order to preserve the isometry), in any coordinate system (where the metric can be written in the above mentioned  form), the Riemann tensors do not have components along the large sphere directions. 
Hence, the contractions of the Riemann tensor do not produce extra orders of $D$, i.e. 
\begin{equation}
	R_{\mu\nu}=\mathcal{O}(D^0)\quad \text{and,}\quad g^{MN}R_{MN}=\mathcal{O}(D^0).
\end{equation}

So, the Einstein tensor part of the Einstein equations for the asymptotic metric is always $\mathcal{O}(D^0)$ and hence the stress tensor components are also $\mathcal{O}(D^0)$. More precisely all components of the internal curvature have the same maximal order in $D$ as the stress tensor. 

We also need to implement the constraints on the dilaton fields mentioned above. We know that the range of values that $e^\phi$ can take is in $[0,\infty)$, hence we can safely do a variable transformation so that $$\phi=2\ln(\psi).$$
The  constraints on the dilaton field in terms of this new variable become
\begin{eqnarray}
	\partial_\alpha \psi \partial^\alpha \psi=1\quad \text{and,}\quad \bar{\nabla}_\alpha\bar{\nabla}_\beta\psi=0.
\end{eqnarray}	
If we write $k_\alpha=\bar{\nabla}_\alpha\psi$, then the above equations become the following constraint on $k_\alpha$,
$$k\cdot k=1 \quad \text{and,} \quad \bar{\nabla}_\alpha k_\beta +\bar{\nabla}_\beta k_\alpha=0,$$
i.e. the radius of the isometry direction is $\psi$ and the effective spacetime contains a unit normalised space-like killing vector along $\big(\partial^\alpha\psi\big)\partial_\alpha$.  We can without loss of generality work in a coordinates system in which radius of the isometry sphere is one of the coordinate direction in effective spacetime and that the metric along the effective spacetime does not depend on this coordinate along $d\psi$. This makes sure that the second of the above two conditions is satisfied. Also, from the fact that $d\psi$ is a unit vector, we infer that $g^{\psi\psi}=1$ in this coordinate system.  Hence, the form of the full spacetime metric is given by
\begin{equation}
	ds^2= g_{\mu\nu}(x^i,\psi)dx^\mu dx^\nu+\psi^2 (x^i)d\Omega^2_{D-p-3}.
\end{equation}
It is easy to see from the structure of the above metric that divergence of any tensor to leading order in large $D$ is in general of the order of $\mathcal{O}(D)$ and this leading order piece is proportional to the component of the tensor along $d\psi$ itself. e.g for vectors 
$$\nabla\cdot V=\frac{1}{\sqrt{\mathcal{G}}}\partial_M\left(\sqrt{\mathcal{G}} \mathcal{G}^{MN} V_N\right)=(D-p-3)\frac{V.d\psi}{\psi}+\mathcal{O}(D^0).$$
This allows for a possibility that even in the Riemann normal coordinates the Taylor series expansion of the metric that we should be working with should contain one extra term than what naive matching with derivative orders would suggest 
\begin{equation}
	g_{MN}=\eta_{MN}+\frac{\epsilon^2}{3} R_{MANB}  y^A y^B+\frac{\epsilon^3}{6}\nabla_C R_{MANB}y^C y^A y^B+\ldots
\end{equation}
Using the above expression of the metric it can be shown that Einstein tensor will contain terms proportional to $\epsilon^3\nabla^C R_{CMNP}|_{x^0} y^N$ which because of the property of divergences in large $D$ is actually $\mathcal{O}(\epsilon^2)$ and has components along $d\psi$. Since, the gravity equations have no other terms proportional to $y^A$ locally ( as the stress tensor is effectively constant at this order), these terms with components along $d\psi$ of the Riemann tensor must be set to zero. This has an interesting implication on the form of the effective spacetime in this region. Namely the components of the metric along $d\psi$ must be equal to their values in the flat spacetime regime and hence in suitably chosen coordinate system the form of the effective spacetime metric must look like
\begin{equation}
	g_{\mu\nu} dx^\mu dx^\nu= g_{ij}(x^i) dx^i dx^j + C d\psi^2 ,
\end{equation}
where, we have used a coordinate system which is $x^\mu\equiv{x^i,\psi}$. Also, since $g^{\psi\psi}=1$ we have $C=1$.  So, the effective metric in a suitable coordinate system has the form
\begin{equation}
	g_{ij}(x^i) dx^i dx^j +  d\psi^2 ,
\end{equation}
where, $\psi$ represents the radius of the large sphere.

  \bibliographystyle{JHEP}
\bibliography{ssbib.bib}
\end{document}